\documentclass[a4paper,11pt,notitlepage]{article}
\usepackage{a4wide,epsfig,graphics,amsthm,amsfonts,amssymb,amsopn,amsmath,longtable,dcolumn,calc,verbatim,threeparttable,eurosym,pdflscape,rotating,multirow, color, ulem, threeparttablex, environ}
\usepackage[paper=a4paper,left=25mm,right=30mm,top=31mm,bottom=31mm]{geometry}

\usepackage[utf8]{inputenc}

\usepackage{natbib}

\usepackage[colorlinks=true,citecolor=black, linkcolor=black, urlcolor=black]{hyperref}

\usepackage{url}
\usepackage{hyperref}

\usepackage{caption}
\usepackage[onehalfspacing]{setspace}
\newcolumntype{.}{D{.}{.}{-1}}
\newcolumntype{d}[1]{D{.}{.}{#1}}

\renewcommand{\baselinestretch}{1.5} 
\theoremstyle{definition}

\theoremstyle{plain}

\newcommand{\bi}{\begin{itemize}}
\newcommand{\ei}{\end{itemize}}

\newcommand{\bc}{\begin{center}}
\newcommand{\ec}{\end{center}}
\newcommand{\bs}{\begin{scriptsize}}
\newcommand{\es}{\end{scriptsize}}
\newcommand{\beq}{\begin{equation}}
\newcommand{\eeq}{\end{equation}}
\newcommand{\ben}{\begin{enumerate}}
\newcommand{\een}{\end{enumerate}}
\newcommand{\bean}{\begin{eqnarray}}
\newcommand{\eean}{\end{eqnarray}}

\DeclareMathSymbol{\R}{\mathalpha}{AMSb}{"52}
\DeclareMathSymbol{\E}{\mathalpha}{AMSb}{"45}

\newcommand{\fe}{\renewcommand{\baselinestretch}{1.32}}

\usepackage{amssymb}
\usepackage{pifont}
\usepackage{float}

\usepackage{tabularx}

\newcolumntype{L}[1]{>{\raggedright\let\newline\\\arraybackslash\hspace{0pt}}m{#1}}
\newcolumntype{C}[1]{>{\centering\let\newline\\\arraybackslash\hspace{0pt}}m{#1}}
\newcolumntype{R}[1]{>{\raggedleft\let\newline\\\arraybackslash\hspace{0pt}}m{#1}}

\usepackage{ifthen}
\makeatletter
\providecommand{\rowno}[1][__empty__]{%
\ifthenelse{\isundefined{\c@rowno}}{%
\newcounter{rowno}}{}%
\ifthenelse{\equal{#1}{__empty__}}{%
\refstepcounter{rowno}%
}{%
\setcounter{rowno}{#1}%
}%
\therowno%
}
\makeatother

\usepackage{subcaption}
\usepackage{graphicx}
\usepackage{caption}
\usepackage{ragged2e,array,longtable}
\usepackage{arydshln}
\usepackage{booktabs}
\usepackage{longtable}

\def\sym#1{\ifmmode^{#1}\else\(^{#1}\)\fi}

\usepackage{xargs}   
\usepackage[pdftex,dvipsnames]{xcolor}  

\begin{document}

\renewcommand{\baselinestretch}{1.12}

\title{\textbf{Long-Term Employment Effects of the Minimum Wage in Germany: New Data and Estimators}}

\author{\textbf{Marco Caliendo}\thanks{%
		e-mail: \texttt{caliendo@uni-potsdam.de}. Corresponding address: University
		of Potsdam, Chair of Empirical Economics, August-Bebel-Str.\ 89,
		14482 Potsdam, Germany. Tel: +49 331 977 3225. Fax: +49 331 977 3210.} \\
	\textit{University of Potsdam, CEPA, IZA, BSE, DIW, IAB} \\
	\textbf{Nico Pestel}\thanks{e-mail: \texttt{n.pestel@maastrichtuniversity.nl}.} \\
	\textit{Maastricht University, IZA, CESifo} \\
	\textbf{Rebecca Olthaus}\thanks{e-mail: \texttt{olthaus@uni-potsdam.de}. 
		\newline The authors thank the Federal Statistical Office and the Statistical Offices of the Federal States for their help in supplying the data. 
  Financial support of the German Minimum Wage Commission for part of this research is gratefully acknowledged.} \\
	\textit{University of Potsdam, CEPA, BSE, DIW}
}


\date{Working Paper\\ \today \vspace{-5mm}}

\maketitle

\begin{abstract}
\noindent 

\noindent We study the long-term effects of the 2015 German minimum wage introduction and its subsequent increases on regional employment. Using data from two waves of the Structure of Earnings Survey allows us to estimate models that account for changes in the minimum wage bite over time. While the introduction mainly affected the labour market in East Germany, the raises are also increasingly affecting low-wage regions in West Germany, such that around one third of regions have changed their (binary) treatment status over time. We apply different specifications and extensions of the classic difference-in-differences approach as well as a set of new estimators that enables for unbiased effect estimation with a staggered treatment adoption and heterogeneous treatment effects. Our results indicate a small negative effect on dependent employment of 0.5 percent, no significant effect on employment subject to social security contributions, and a significant negative effect of about 2.4 percent on marginal employment until the first quarter of 2022. The extended specifications suggest additional effects of the minimum wage increases, as well as stronger negative effects on total dependent and marginal employment for those regions that were strongly affected by the minimum wage in 2015 and 2019. 

\vspace{0.3cm}

\end{abstract}

\fe

\thispagestyle{empty}

\newpage

\setcounter{section}{0}
\setcounter{page}{1}
\pagenumbering{arabic}
    \section{Introduction}

The introduction of a statutory minimum wage in Germany in 2015 was one of the largest labour market policy reforms in many years, and it had been controversially debated in both academia and politics. Against the background of the minimum wage debate in the literature since the early-1990s \citep{CardKrueger1994, CardKrueger2000, NeumarkWascher2000} and the controversy about the direction of employment effects of minimum wages \citep{NeumarkEtAl2014, AllegrettoEtAl2017}, the contemporary discussion at the time focused on the question of whether the German minimum wage of initially \euro{8.50} per hour would cause substantial employment losses.\footnote{Upon introduction, the minimum wage equalled about 49.8\% (57.6\%) of the mean (median) wage in Germany and about 11.4\% of workers were affected by it (Structure of Earnings Survey, 2014).}  

In the meantime, the short- to medium-term employment effects of the German minimum wage introduction have been extensively studied in the literature \citep[for summaries see][]{CaliendoEtAl2018, CaliendoEtAl2019, Bruttel2019, BosslerGerner2020, Mindestlohnkommission2020}.\footnote{There also exist older studies about the previously introduced sector-specific minimum wages, mostly not finding significant employment effects (see \citeauthor{BoschWeinkopf2012}, \citeyear{BoschWeinkopf2012} and \citeauthor{Moeller2012}, \citeyear{Moeller2012} for a detailed overview and \citeauthor{DoerrFitzenberger2016}, \citeyear{DoerrFitzenberger2016} for a critical assessment of the applied methods). Nonetheless, there were some signs of a tendency towards negative labour market effects in East Germany, where the impact intensity of the minimum wage is substantially higher than in West Germany due to the overall lower wage level.} Overall, most studies find no or only small negative effects of the minimum wage introduction on total employment \citep{Mindestlohnkommission2018a, BoninEtAl2018, CaliendoEtAl2019, BruttelEtAl2019, PestelEtAl2020}. 
Hereby, the effect is predominantly due to marginal employment, while employment subject to social security contributions remained stable \citep{CaliendoEtAl2018, BoninEtAl2018, AhlfeldtEtAl2018, Garloff2019, HoltemoellerPohle2020, Schmitz2019, PestelEtAl2020}.\footnote{\cite{Friedrich2020} even identifies positive effects of the minimum wage on employment subject to social security contributions until the first increase in 2017.
\cite{PestelEtAl2020} find that the effects can be mainly attributed to the introduction of the minimum wage, while the first increase had no considerable additional effect, and regions/sectors with a below-average growth dynamic experienced stronger negative employment effects.   
\cite{Bossler2016, Bossler2017} and \cite{BosslerGerner2020} show that the minimum wage had an impact on the employment dynamic by reducing the number of new hires. 
While focusing mainly on effects on wage inequality induced by the minimum wage, \cite{BosslerSchank2023} show that their finding of a significantly reduced inequality in monthly earnings is not driven by employment effects as they find no significant effects on employment. \cite{BiewenEtAl2023} also identify a significant increase in wages and a reduction in wage inequality due to the minimum wage introduction, but no significant effect on the distribution of weekly working hours.    
\cite {DustmannEtAl2022} find that instead of the feared employment effects, a reallocation of the employees subject to the minimum wage into more productive firms took place. Hereby, they identify substantial wage gains in the lower part of the wage distribution, of which one quarter can be explained by changes from smaller (relatively low paying) to larger (relatively high paying) firms. This ``upgrade'' only occurs for low-wage workers and not for higher-earning workers.}

The existing literature mostly focuses on the short-term effects two to three years following the minimum wage introduction, while much less is known about the medium- and especially long-term effects. This is relevant for different reasons. First, employment adjustments may unfold over a longer period than simply a few years, especially in the setting of the German labour market where firing is rather difficult. Second, the minimum wage has been increased multiple times and may have caused additional employment effects after the initial adjustment phase of the introduction. In 2017 and 2019, the minimum wage was raised to \euro{8.84} and \euro{9.19}, respectively, and since then it has been increased several more times to a current level of \euro{12}.\footnote{See Table \ref{tab:mw_development} in the Appendix for an overview of the development of the minimum wage level over time.} Additionally, previous studies are mainly based on data from the 2014 (pre-introduction) Structure of Earnings Survey (SES) to determine the treatment status across labour market regions in Germany. However, the composition of regions mainly affected by the introduction may differ from that of subsequent major increases of the minimum wage.

This paper studies the longer-run employment effects of the minimum wage introduction in Germany. The analysis is based on aggregate administrative data on employment at the level of regional labour markets over the period from 2013 to 2022. Regions were differently affected by the minimum wage introduction due to regional differences in wage levels and the distribution of wages. Thus, there is a strong variation in the regional minimum wage bite. We exploit this variation to apply a difference-in-differences (DiD) approach to estimate the causal effects of the minimum wage on employment. We operationalise the regional minimum wage bite using the regional wage gap based on the SES 2014. Given that the SES is conducted every four years, we can measure the regional wage gap again in 2018 with respect to the minimum wage level after it had been substantially raised. Thus, we have a measure of the degree of regional exposure to the minimum wage at two different points in time. Overall, the wage gap substantially decreased between 2014 and 2018, while the fraction of affected workers declined to a much lesser extent. We observe that a substantial share of German labour market regions changed between treatment and control status between 2014 and 2018. To the best of our knowledge, we are the first to use two waves of the SES to study the employment effects of the minimum wage in Germany.

In our main specification, we estimate the effects of the minimum wage introduction alone using a treatment indicator based on the SES 2014. In a second step, we account for the increases of the minimum wage and the substantial change in strongly affected regions over time in multiple ways. First, we add interaction terms for the minimum wage increases to the basic specification to check for additional effects of the raises. Second, we apply a specification with three treatment groups that are defined by having a relatively high minimum wage bite in 2014, 2018 or both years. Regions with a relatively low minimum wage bite in both years form the control group. Finally, we treat the minimum wage introduction and the increase in 2019 as one treatment that was introduced in a staggered manner, where some regions start treatment in 2014 and others in 2019. In this setting, the classic TWFE estimator may be biased if treatment effects are heterogeneous across groups/units or time. This bias can be caused by so-called ``forbidden comparisons'' \citep{BorusyakJaravel2017} between the outcomes of later- and earlier-treated units.
  
Several different new estimators have recently been developed to obtain unbiased DiD estimates despite a setting deviating from the canonical one with two groups and two periods. We compare the estimates of the classic TWFE with those of new estimators for the longer-term effects of the German minimum wage on employment if treatment adoption is considered to be staggered. We apply the estimators developed by \cite{SunAbraham2021}, \cite{CallawaySant'Anna2021}, \cite{BorusyakEtAl2021}, and \cite{dCh_dH2022b}.

Our main findings confirm the negative but moderate employment effects of the minimum wage introduction, which is mainly driven by significant reductions in marginal employment. While the effect on total dependent employment has decreased in magnitude over time, the effect has amplified over time for marginal employment. The negative employment effect on marginal employment is also substantially stronger for regions with a relatively low GDP growth rate prior to the introduction of the minimum wage. We find that the regional minimum wage bite decreased in all regions, albeit not at the same rate, whereby the ranks of the regional bite substantially changed between 2014 and 2018. One consequence is that while the initial minimum wage introduction mainly affected the labour market in East Germany, the recent minimum wage hikes are also increasingly affecting lower-wage regions in West Germany. This results in the finding that the minimum wage increases -- especially the one in 2019 -- have resulted in additional employment effects on top of the longer-term effects of the minimum wage introduction of 2015. Regions with a relatively high wage gap in 2014 and 2018 experience substantially stronger negative employment effects than those that only have a high wage gap in one of the two years. Finally, we document that the alternative DiD estimators overall support our main findings. 

The remainder of the paper is structured as follows. Section \ref{sec:data_desc} describes the data and provides descriptive results. In Sections \ref{sec:method} and \ref{subsec:results}, we explain the empirical approach and present our regression results. 
Section \ref{sec:conclusion} concludes.

    \section{Data and Descriptive Results} \label{sec:data_desc}

\subsection{Data Sources}\label{subsec:data_source}

We use the SES of 2014 and 2018 to measure the regional minimum wage bite.\footnote{Source: RDC of the Federal Statistical Office and Statistical Offices of the Federal States, Structure of Earnings Survey, 2014 (DOI:10.21242/62111.2014.00.00.1.1.1) and 2018 (DOI:10.21242/62111.2018.00.00.1.1.0).} The SES is conducted every four years and the data is provided by the Federal Statistical Office of Germany. It includes information on earnings of employees in Germany as well as other detailed information about employment, such as the hours worked, industry, and personal characteristics of the employees. The data covers the month of April of the respective year and thus stems from a point in time directly prior to the introduction of the minimum wage, which was announced in July 2014. Firms are obligated to provide information and for a high level of representativeness, firms are chosen stratified for states, industry and firm size category \citep{FDZ2019}. The SES 2014 (2018) contains information about approximately one million employment relationships from about 71,000 (60,000) companies with at least one employee subject to social security contributions. The SES 2014 has previously been used for evaluating the minimum wage's employment effects \citep{BruttelEtAl2018, BoninEtAl2018, CaliendoEtAl2018, PestelEtAl2020} although -- to our knowledge -- we are the first to use the SES 2018. 

The SES has two major advantages compared to other data sources that are firm-based: first, it allows a precise estimation of hourly wages on the individual level, which in turn allows determining the bite on different dimensions; and second, the large number of observations enables an analysis at the level of the 257 labour market regions, as in our analysis. However, results below the regional level of the federal state are not representative per se \citep{FDZ2019}. Another caveat is that the SES does not include regional information for civil servants below the level of the federal states, and thus we have to exclude them from our analysis. 
Since the SES 2014 and 2018 share the same characteristics, the SES 2018 allows smoothly continuing with the minimum wage evaluation and showing developments between the year prior to the minimum wage introduction and the year prior to the second increase in 2019. The 257 labour market regions are divided into a control and treatment group based on the wage gap calculated using the SES 2014. We also divide the labour market regions into a treatment and control group based on the SES 2018, which documents the development of the degree of exposure over time and which we will use in additional analyses. 

The data for our outcome measures come from the statistical department of the Federal Employment Agency, which publishes the number of employed people on a quarterly basis at a regional level. The statistics are based on process data of the Federal Employment Agency as well as the statutory pension insurance and can be considered as very reliable.  
In our analysis, we use data from the first quarter of 2013 until the first quarter of 2022. 
The Federal Statistical Office differentiates between marginal employment and employment subject to social security contributions, and further between exclusive marginal employment and marginal employment as a side job. We calculate the total number of persons in dependent employment as the sum of employment subject to social security contributions and exclusive marginal employment.
Further, we use data for regional economic and demographic indicators such as GDP and population from the regional statistics of the Federal Statistical Office. Moreover, we include regional information about the settlement structure (classification in urban and rural areas) from the Federal Institute for Research on Building, Urban Affairs and Spatial Development.


\subsection{Treatment Indicator}\label{subsec:treatment}

In order to identify a causal effect of the minimum wage on employment outcomes in Germany, we use a DiD approach (see Section \ref{sec:method}).
However, due to the universal nature of the minimum wage in Germany, there is no suitable control group that is not subject to the minimum wage.\footnote{There were a few exceptions from the minimum wage upon its implementation in 2015, yet the number of excepted employees is quite low, and these groups cannot be adequately compared to regular employees as they are mostly apprentices, interns, and teenagers below the age of 18. Further, sectors with their own minimum wages above the general minimum wage are also not an appropriate control group, because they represent a very selective sample of the labour market.} Therefore, we divide regions into a control and a treatment group based on the degree to which they are affected by the minimum wage. To measure the degree to which a region is affected, we use the wage gap, describing the average absolute difference between the hourly wage and the minimum wage of \euro{8.50} for the SES 2014 (\euro{9.19} for the SES 2018) for hourly wages below the minimum wage, while the difference is equal to zero for wages of at least \euro{8.50} (\euro{9.19}):

	\begin{equation*}
		WageGap_{it} = \frac{\sum_{w=1}^{N_{it}} max[MW_{t} - Wage_{wit} , 0]}{N_{it}}, 
	\end{equation*}

with $i$ representing the region, $t$ the year (2014 or 2018), $w$ the individual worker, and $N$ the number of workers. The concept of the wage gap that we use is based on the population of all employees not exempt from the minimum wage and quantifies the minimum wage exposure in respect to the number of employees as well as considering the level of the minimum wage relative to the base level of wages. We focus on the wage gap, because -- in contrast to other measures (i.e. Kaitz- or Fraction-Bite) -- it quantifies the size of the adaption of wages that is necessary to raise the hourly wage from April 2014 (April 2018) to the level of the minimum wage of \euro{8.50} (\euro{9.19}). 
The final division between the treatment and control group is contingent on the median of the regional wage gaps (weighted by the regional population in 2013). Regions with a minimum wage gap above (below) the weighted median are part of the treatment (control) group.


\subsection{Descriptive Statistics}\label{subsec:descriptives}

\begin{figure}[h!] \caption{Regional wage gap to \euro{8.50} in 2014 (left) and to \euro{9.19} in 2018 (right)} \label{fig:map_gaps}
\center	
\label{fig:wagegap_maps}\includegraphics[scale=0.35]{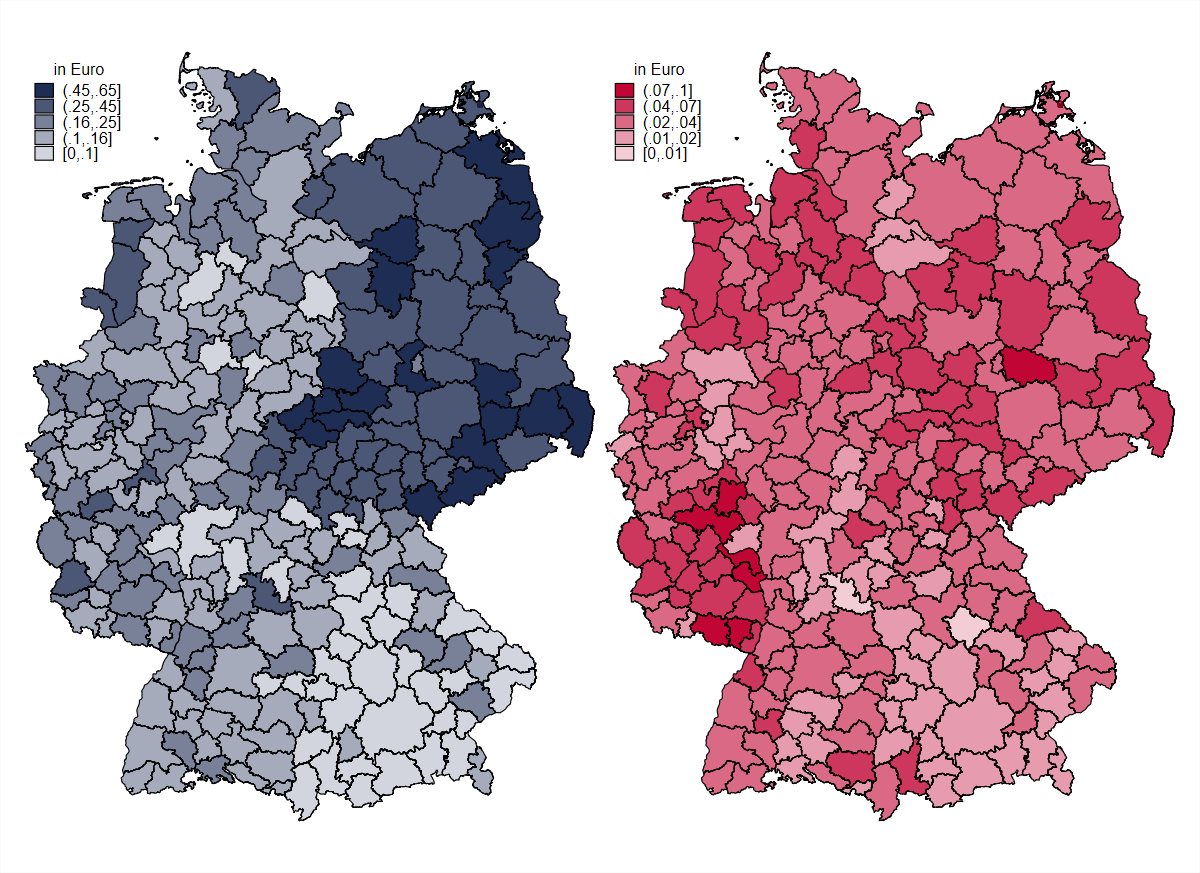}
	\caption*{\footnotesize Source: SES 2014, 2018; own calculations. Note: The map on the left shows the regional wage gap to \euro{8.50} in 2014. The map on the right shows the regional wage gap to \euro{9.19} in 2018. The respective wage gaps are grouped into five categories (separately for 2014 and 2018) with a different shade of blue (2014) or red (2018) for each of them. For the definition of the wage gap, see Section \ref{subsec:treatment}.}
\end{figure}

\paragraph{Minimum Wage Bite Over Time}
The map on the left-hand side of Figure \ref{fig:wagegap_maps} shows the wage gap for all labour market regions in 2014. The wage gap is especially large in East Germany and lowest in the south, especially in the federal state of Bavaria. The size of the wage gap varies between 0.021 and 0.652. The right-hand side panel of Figure \ref{fig:wagegap_maps} shows the wage gap for all labour market regions in 2018. While the wage gap has declined in all regions, the geographical distribution is no longer as clear. Especially, East Germany does not stand out any longer and there are substantially more regions in West Germany with an above-median wage gap. This change also becomes clear when looking at the correlations of the wage gaps in 2014 and 2018. The Pearson's correlation coefficient with 0.295 and the Spearman rank correlation coefficient of 0.4082 show that the two wage gaps are only weakly correlated.\footnote{For illustrative purposes, Figure \ref{fig:scatter_ranks} in the Appendix shows scatter plots for the wage gap ranks 2014 (8.50\euro) / 2018 (9.19\euro), 2014 (8.50\euro) / 2014 (9.19\euro), and 2014 (9.19\euro) / 2018 (9.19\euro). The wage gap ranks of 2014 and 2018 show no strong relationship (irrespective of whether the wage gap in 2014 is relative to 8.50\euro \hspace{0.1cm} or 9.19\euro) while the two wage gaps in 2014 are very closely and positively related. This suggests that the change in wage gap ranks between 2014 and 2018 is not due to differences in the wage distribution in an interval between \euro{8.50} and \euro{9.19} that already existed in 2014.}

Table \ref{tab:desc_main} in the Appendix takes a closer look at characteristics of the labour market regions in 2014 and 2018 for the whole group as well as those with a relatively low and high wage gap separately. In 2014, 144 labour market regions belonged to the treatment group and 113 regions to the control group. The average overall wage gap was 0.203, while it was 0.281 in the treatment group and 0.104 in the control group. In 2018, the number of regions in the treatment group increased to 153, while the number in the control group decreased to 104. With 0.034, the average wage gap was substantially lower in 2018 than in 2014. Hereby, the average was 0.043 in the treatment and 0.019 in the control group.
The settlement structure of regions in the two groups only slightly varied between 2014 and 2018. In both years, there are more urban regions in the control group and more rural areas in the treatment group.
Further, there are no large differences between the treatment and control group or across years regarding the employment structure by sectors and in terms of the population share between the ages of 18 and 64 in 2013. In both years, the labour market regions in the treatment group experience a lower GDP growth between 2010 and 2013 and between 2015 and 2018. The share of regions with a GDP growth rate between 2010 and 2013 in the bottom quartile is higher in the control group than the treatment group in 2014. However, in 2018, the share is almost equal. The share of regions with a GDP growth rate between 2015 and 2018 in the bottom quartile is substantially higher in the treatment groups in 2014 and 2018.

 \begin{figure}[b!] \caption{Treatment groups based on the wage gaps in 2014 and 2018} \label{fig:map_groups}
\center
\label{fig:geb_main}\includegraphics[scale=0.18]{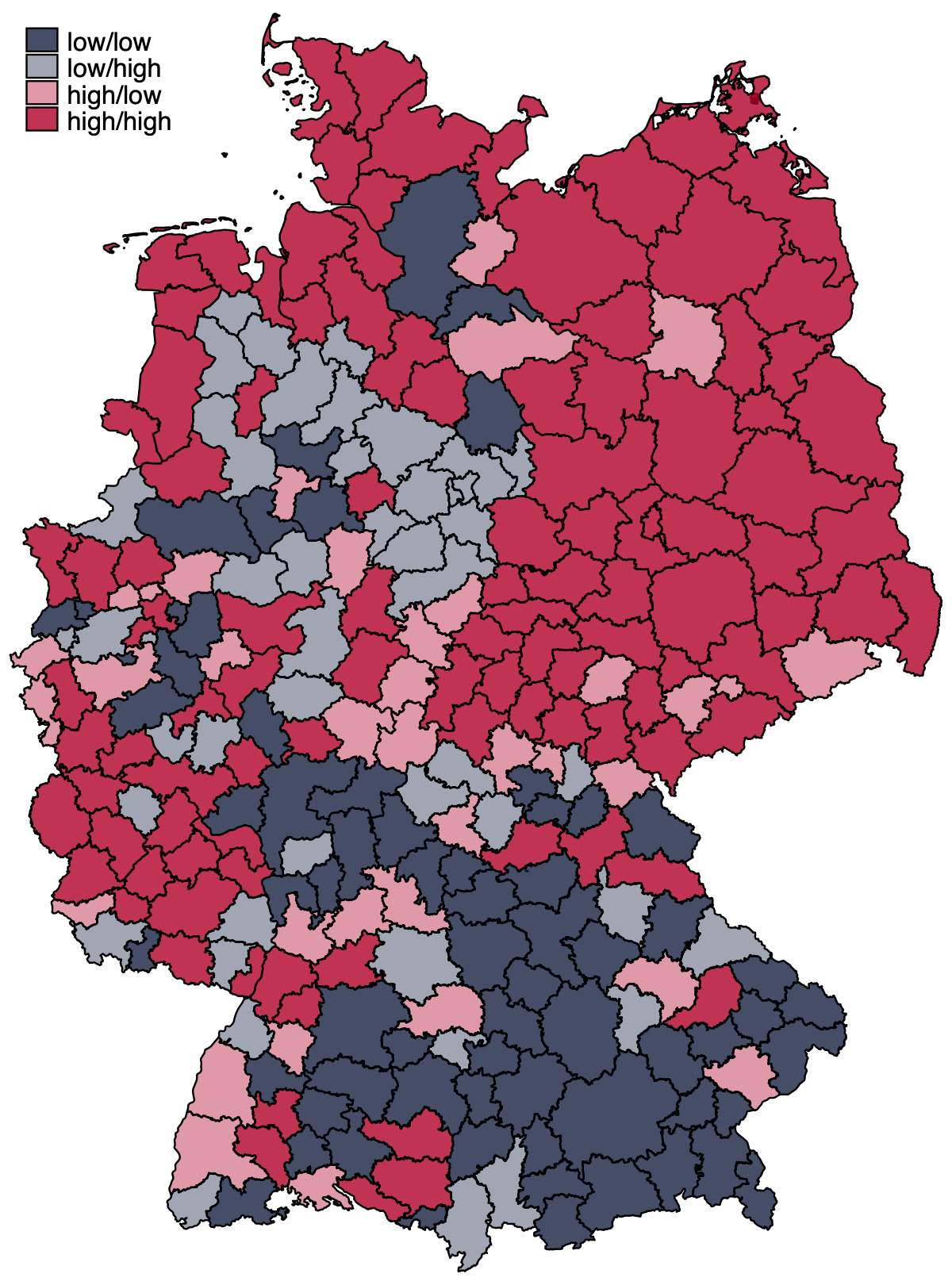}
	\caption*{\footnotesize Source: SES 2014, 2018; own calculations. Note: The map shows the labour market regions sorted into four groups based on their wage gaps in 2014 and 2018. ``low'' describes a wage gap below the weighted median of all regions in 2014 or 2018 and ``high'' describes a wage gap at or above that median. For the definition of the wage gap, see Section \ref{subsec:treatment}.}
\end{figure}

\paragraph{Switching Treatment Status} 

The majority of labour market regions do not change their binary treatment status between 2014 and 2018. Sixty-eight regions have a relatively low wage gap in both years and 108 regions a relatively high wage gap. However, a substantial share of about 31.5 percent of the regions that change between a low and high wage gap in that period. Forty-five regions switched from a relatively low to a relatively high wage gap and 36 regions did the opposite. Based on this observation, we categorise the labour market regions into four different groups: ``low/low group'' (relatively low wage gap both in 2014 and 2018), ``low/high group'' (a low wage gap in 2014 and a high wage gap in 2018), ``high/low group'' (a high wage gap in 2014 and a low wage gap in 2018), and ``high/high group'' (a high wage gap in both years). We will use these four groups in one of our DiD specifications described in Section \ref{sec:method}, where we estimate separate treatment effects for each group.  

Figure \ref{fig:map_groups} is a map showing the regional distribution of these four groups. While most East German regions are in the high/high group and many southern regions belong to the low/low group, there are especially (but not exclusively) many regions in the state of Lower Saxony in north-western Germany that switched between a relatively high and low wage gap. Table \ref{tab:desc_groups} in the Appendix shows descriptive statistics similar to Table \ref{tab:desc_main} for the four groups. By definition, the average wage gap in 2014 (2018) is larger in the high/low and high/high (low/high and high/high) groups. Of special interest are the switchers. The regions in the low/high group decrease their minimum wage exposure in 2018 on average by almost 63 percent compared to 2014. At the same time, the regions in the high/low group reduced their exposure between 2014 and 2018 by about 91 percent. All groups reduced the fraction of employees directly affected by the minimum wage, but again there is large variation between the groups. The fraction in the low/high group only decreased by about 6 percent, while the high/low group experienced a large decrease in affected employees of approximately 62 percent. The average hourly wage was significantly lower for the high/low group than for the low/high group in 2014, although in 2018 the averages are virtually the same across both groups. Accordingly, the hourly wages increased at more than twice the rate in the high/low group compared with the low/high group (17.8 to 8.0 percent). These numbers suggest that the development of wages following the introduction of the minimum wage varied between groups and different regional developments between 2014 and 2018 led to changes in the exposure to the minimum wage increase relative to its introduction.

\paragraph{Development of Outcomes Over Time} Figure \ref{fig:outcomes} in the Appendix shows the development of total employment over time separately for labour market regions with a high and a low wage gap upon its introduction. During the observation period, the dependent employment generally increases (with small seasonal cycles). It increased slightly more strongly in regions with a low wage gap and has been at the same level in both groups since the end of 2018. The employment subject to social security contributions makes up the large majority of the dependent employment and hence moves very similarly over time. 

Marginal employment remained stable until the end of 2019 in regions with a low wage gap, while it decreased in regions in the treatment group after the introduction of the minimum wage. This already hints towards a reduction in marginal employment in the treatment group through the introduction of the minimum wage. The difference between the two groups increased over time and has remained constant from 2019 onwards. Throughout the Covid-19 pandemic, marginal employment strongly decreased in both groups, but the difference between them remained almost unchanged. 

    \section{Empirical Approach} \label{sec:method}

Our baseline DiD regression equation for estimating the cumulative effect of the German minimum wage on employment outcomes over the observation period reads as follows:

\begin{equation} \label{eq:equation1}
	Log(Y_{it}) = \beta(Wage Gap_{i,2014}^{high} \times I_{t>Q2/2014}) + X_{it}\gamma + \theta_{i} + \theta_{t} + \epsilon_{it},
\end{equation}

\noindent where $Log(Y_{it})$ is the log of the dependent variable in labour market region $i$ in quarter $t$. $\theta_{i}$ represents regional fixed effects that control for all time-invariant characteristics such as geographic location. $\theta_{t}$ adds quarter fixed effects that capture time-specific effects such as the overall economic development. The treatment variable in our DiD analysis $Wage Gap_{i,2014}^{high} \times I_{t>Q2/2014}$ is the interaction between a binary indicator for regions with a wage gap at or above the median and a binary indicator for post-treatment quarters. The treatment period in our specification starts after the second quarter of 2014 to account for potential anticipation effects since the law introducing the minimum wage was passed in July 2014. Consequently, $\beta$ is the coefficient of interest that measures the average treatment effect of a relatively high minimum wage bite on the respective outcome. $X_{it}$ is a vector of control variables that can vary between labour market regions and over time such as differing trends for urban and rural regions.\footnote{The control variables included are interactions of: time, east, and population share between 18 and 64 years in 2013; time, east, and type of labour market region; time, east, and employment share in different sectors; and time and GDP per capita in 2013.} We cluster the standard errors at the level of the labour market regions to allow for correlation of unobservable characteristics of a labour market region over time. Further, observations are weighted with the number of employees in April 2014, such that the results are not driven by relatively small labour market regions. 

To analyse the dynamic development of the minimum wage effects over time, we apply an event-study approach. The corresponding estimation equation reads as follows:
\begin{equation} \label{eq:equation2}
	Log(Y_{it}) = \sum_{\tau = Q1/2013,\tau \neq Q2/2014}^{Q3/2021} \beta_{\tau}(Wage Gap_{i,2014}^{high} \times I_{t = \tau})+X_{it}\gamma + \theta_{i} + \theta_{t} + \epsilon_{it},
\end{equation}

\noindent where $Wage Gap_{i,2014}^{high}$ is interacted with indicators for all quarters except for the baseline period (the second quarter of 2014). Thus, the coefficient vector $\beta_{\tau}$ contains the estimated treatment effect for every quarter before and after the minimum wage was announced relative to the baseline quarter. Insignificant estimates for pre-treatment periods support the identifying assumption of parallel trends between treated and control regions. 

We also estimate Equations (\ref{eq:equation1}) and (\ref{eq:equation2}) with an additional interaction of the treatment indicator with a binary indicator for a relatively low GDP growth between 2010 and 2013 to test whether the treatment effect varies with the regional growth dynamic prior to the introduction of the minimum wage. Hereby, ``low'' is defined as a regional GDP growth rate in the bottom quartile of all regions. 

We also extend Equation (\ref{eq:equation1}) to study additional effects of the minimum wage raises on January 1 of 2017 and 2019 to 2022: 

\begin{equation} \label{eq:equation3}
\begin{split}
	Log(Y_{it}) = \beta(Wage Gap_{i,2014}^{high} \times I_{t>Q2/2014}) + \sum_{\tau = 2016, \tau \neq 2017}^{2021} \delta_{\tau}(Wage Gap_{i,2014}^{high} \times I_{t>Q4/\tau}) \\
 + X_{it}\gamma + \theta_{i} + \theta_{t} + \epsilon_{it},
 \end{split}
\end{equation}

\noindent where the coefficients $\delta_{\tau}$ indicate the additional effects of minimum wage raises beyond the initial introduction in 2015. Additionally, we also apply a shortened version only including the first two increases. 

We exploit variation in the minimum wage bite between 2014 and 2018 by estimating an additional regression model using the four groups of regions that we introduced in Section \ref{subsec:descriptives}. These groups are based on the regional wage gaps relative to the respective population-weighted median in 2014 and 2018: low/low, low/high, high/low, and high/high. We include treatment indicators for the last three groups, which makes the first group the control group. The corresponding estimation equation is: 
\begin{equation} \label{eq:equation4}
	\begin{split}	
		Log(Y_{it}) = \beta(Wage Gap_{i,2014,2018}^{low,high} \times I_{t>Q2/2014}) + \gamma(Wage Gap_{i,2014,2018}^{high,low} \times I_{t>Q2/2014}) \\
		+ \delta(Wage Gap_{i,2014,2018}^{high,high} \times I_{t>Q2/2014}) + X_{it}\lambda + \theta_{i} + \theta_{t} + \epsilon_{it}.
	\end{split}
\end{equation}

We also add a placebo term in the form of $\alpha(Wage Gap_{i,2014}^{high} \times I_{t<Q2/2014})$ to Equations (\ref{eq:equation1}) and (\ref{eq:equation3}). Equation (\ref{eq:equation4}) contains one placebo term of this kind for each of the three treatment groups. These placebo terms show whether the trends of the treatment and control group prior to the treatment were significantly different from each other. If this is not the case, it supports the common trend assumption necessary to identify a causal effect.

Finally, we also use an alternative estimation based on a staggered treatment adoption. Here, treatment starts in the third quarter of 2014 for those regions with a relatively high wage gap in 2014 (as is the case in the baseline setting) and in 2019 for regions that had a relatively low wage gap in 2014 and a high wage gap in 2018. Regions with a relatively low wage gap in both periods are never treated in this setting. 
The estimation equation reads as follows: 
	\begin{equation} \label{eq:equation5}
 \begin{split}
	Log(Y_{it}) = \beta(Wage Gap_{i,year}^{high} \times I_{t\geq F}) +  X_{it}\gamma + \theta_{i} + \theta_{t} + \epsilon_{it},
 \end{split}
	\end{equation}

with $F$ referring to the time when a region starts treatment for the first time (Q3 2014 or Q1 2019). The corresponding year of the binary minimum wage bite is either 2014 or 2018. Once a region is treated, it remains treated for the whole observation period.  
Recent advances in the econometrics literature have shown that in settings with a staggered treatment adoption, TWFE estimates can be biased if treatment effects are heterogeneous across groups/units or time \citep[see, for example,][]{SunAbraham2021}. If units are treated at different points in time, estimates of a classic TWFE model can include ``forbidden comparisons'' \citep{BorusyakJaravel2017} between two treated groups or units (one earlier treated and one later treated). In case of heterogeneous treatment effects, this can cause the TWFE estimates to be biased. We see potential sources for such a treatment heterogeneity; for example, the treatment effect could differ between regions starting treatment in 2014 and 2019 or because the treatment intensity (the continuous wage gap) even differs within the treatment groups. Another possible reason is varying economic growth across regions that could affect the size of the treatment effect. 

Several different new estimators have recently been developed to obtain unbiased DiD estimates despite a treatment that is adopted in a staggered manner. Usually, the idea behind the newly developed estimators is to only include ``clean'' comparisons between observations that are treated and those that have not (yet) been treated. These comparisons are then aggregated using different weights (that are to be chosen).\footnote{See \cite{dCH_dH2022a} and \cite{RothEtAl2023} for comprehensive overviews of the recent advances in the DiD literature.}
We compare the estimates of the classic TWFE with those of some of the newly developed estimators. The estimator developed by \cite{SunAbraham2021} generalises the event-study approach to settings with a staggered treatment adoption and is based on cohort- and period- specific effect estimates. Treatment start is defined as the time when a unit’s treatment status changes for the first time and the control group is either the never-treated or the last-treated group(s). By contrast, the estimator by \cite{CallawaySant'Anna2021} can use the not-yet-treated or never-treated groups as controls. The basis of their approach is the estimation of group-time-average treatment effects. For example, these can be aggregated over groups (a group is defined by a common time of treatment start), by time relative to treatment start, or over all groups and periods to obtain an overall average treatment effect on the treated.
The estimator proposed by \cite{dCh_dH2022b} generalises ``the event-study approach to such designs, by defining the event as the period where a group’s treatment changes for the first time'' (\citeauthor{dCH_dH2022a}, \citeyear{dCH_dH2022a}, p.20).
\cite{BorusyakEtAl2021} apply a different approach than the previously described estimators, developing an imputation estimator (others like \cite{Gardner2022} use similar approaches). First, the counterfactual outcome for the treated observations is predicted by regressing the outcome on group and time fixed effects for the sample of not-treated observations. In a second step, the counterfactual outcome is subtracted from the observed outcome for the treated observations to obtain the treatment effect \citep{BorusyakEtAl2021, dCH_dH2022a}.
    \section{Results}\label{subsec:results}

In this section, we present our regression results. We start in Section \ref{subsec:results_main} by showing the estimates for the employment effects of the minimum wage introduction based on Equations (\ref{eq:equation1}) and (\ref{eq:equation2}).
Additionally, we also estimate the treatment effects if the binary treatment is interacted with regional economic growth prior to the introduction of the minimum wage. In Section \ref{subsec:results_increases}, we present results for the incremental changes of the minimum wage according to Equation (\ref{eq:equation3}). This is followed by estimations with three different treatment groups (Equation (\ref{eq:equation4})) in Section \ref{subsec:results_groups} and results for a staggered treatment adoption in Section \ref{subsec:results_staggered} (Equation (\ref{eq:equation5})).

\subsection{Long-Term Effects of the Minimum Wage Introduction}\label{subsec:results_main}

\paragraph{Baseline results.} 
We present our baseline findings for the long-term effects of the minimum wage introduction on employment outcomes in Table \ref{tab:results_main_bip}. Panel A shows the treatment estimates according to Equation (\ref{eq:equation1}) for dependent employment, employment subject to social security contributions, and marginal employment including all of the covariates described in Section \ref{sec:method}.   

Column (1) of Panel A in Table \ref{tab:results_main_bip} shows that the minimum wage introduction had a significant negative impact on the total dependent employment. The effect size is about $-0.5$ percent and statistically significant at the 10 percent level. The coefficient for the employment subject to social security contributions in column (2) is positive but very small and not statistically significant. Column (3) of Table \ref{tab:results_main_bip} shows the results for marginal employment. The introduction of the minimum wage had a negative and statistically significant effect (at the 1 percent level) on the marginal employment of about $–2.4$ percent. The placebo terms are insignificant for all three outcomes, thereby supporting the assumption of common pre-trends. 

\begin{scriptsize}
	\begin{table}[htbp]
  \centering
  \caption{Effects of the minimum wage introduction on regional employment}
  \label{tab:results_main_bip}
    \begin{tabular}{lccc}
    \toprule
    \toprule
    \multirow{2}{2cm}{VARIABLES} & Dependent & Employment  & Marginal  \\
    & employment & subject to SSC  & employment  \\
    \midrule
     \multicolumn{4}{l}{\textbf{Panel A: Binary treatment (wage gap)}} \\
    \multicolumn{1}{l}{Treatment} & -0.00511* & 0.000456 & -0.0239*** \\
    \multicolumn{1}{l}{} & (0.00260) & (0.00286) & (0.00763) \\
    \multicolumn{1}{l}{Placebo} & 0.000569 & 0.00142 & -0.00140 \\
    \multicolumn{1}{l}{} & (0.00111) & (0.00107) & (0.00246) \\
          &       &       &  \\
    \multicolumn{1}{l}{R$^{2}$} & 0.597 & 0.584 & 0.472 \\
          &       &       &  \\
     \multicolumn{4}{l}{\textbf{Panel B: Binary treatment (wage gap) interacted with growth}} \\
    \multicolumn{1}{l}{Treatment} & -0.00434 & 0.000229 & -0.0192** \\
    \multicolumn{1}{l}{} & (0.00332) & (0.00357) & (0.00888) \\
    \multicolumn{1}{l}{Placebo (Treatment)} & 0.000593 & 0.00172 & -0.00222 \\
          & (0.00124) & (0.00119) & (0.00281) \\
    \multicolumn{1}{l}{Treatment x Low growth 2010-2013} & -0.00571 & -0.00159 & -0.0232* \\
          & (0.00549) & (0.00661) & (0.0121) \\
    \multicolumn{1}{l}{Placebo (Treatment x Low growth} & 0.00152 & 0.000757 & 0.00417 \\
    \multicolumn{1}{l}{2010-2013)} & (0.00228) & (0.00223) & (0.00473) \\
          &       &       &  \\
    \multicolumn{1}{l}{R$^{2}$} &  0.605     &   0.590    & 0.484 \\
    \midrule
    \multicolumn{1}{l}{Observations} & 9,509 & 9,509 & 9,509 \\
    \midrule
    \multicolumn{1}{l}{Labour market region FE} & X     & X     & X \\
    \multicolumn{1}{l}{Quarter FE} & X     & X     & X \\
    \multicolumn{1}{l}{Controls} & X     & X     & X \\
    \bottomrule
    \bottomrule
    \end{tabular}%
  \caption*{\scriptsize Source: Regional Statistic of the Federal Employment Agency; SES 2014; Federal Statistical Office; Federal Institute for Research on Building, Urban Affairs and Spatial Development; own calculations. Note: The treatment effect refers to the coefficient $\beta$ in Equation (\ref{eq:equation1}), estimated with TWFE. The binary treatment equals 1 if the regional wage gap is equal to or above the population-weighted median. 
  The control variables included are interactions of: time, east, and population share between 18 and 64 years in 2013; time, east, and type of labour market region; time, east, and employment share in different sectors; time and GDP per capita in 2013. The specification in Panel B additionally includes the interactions of time and the indicator for a low GDP between 2010 and 2013. Standard errors (in parentheses) are clustered at the level of the labour market regions. Confidence level: ***p$<$0.01, **p$<$0.05, *p$<$0.1.}
\end{table}%

 \end{scriptsize}

\begin{figure}[h!] \caption{Effect of the minimum wage on employment} \label{fig:event_main}
	\begin{subfigure}{0.5\textwidth}
		\subcaption{Dependent employment} \label{fig:svgeb_main}\includegraphics[scale=0.18, trim=0 0 0 0, clip]{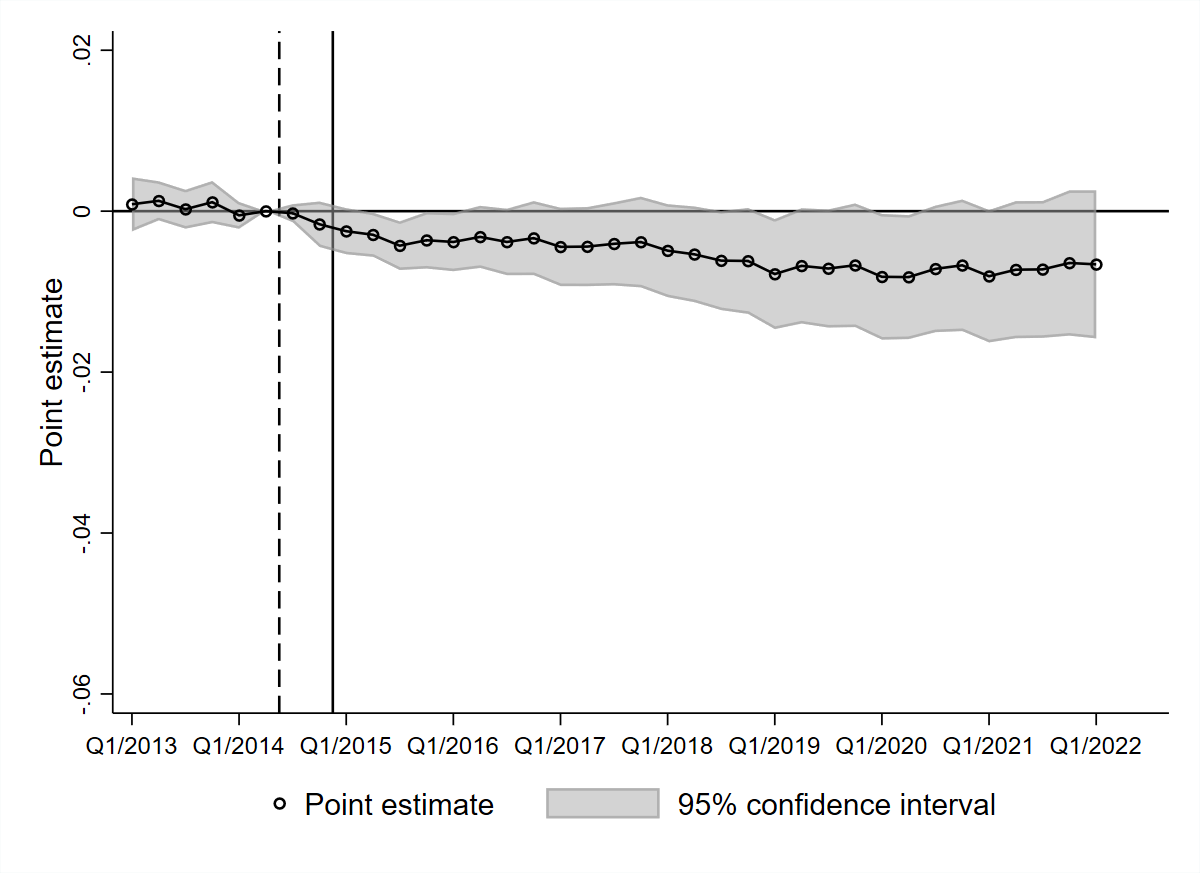}
	\end{subfigure}
	\begin{subfigure}{0.5\textwidth} 
		\subcaption{Employment subject to SSC} \label{fig:svb_main}\includegraphics[scale=0.18, trim=0 0 0 0, clip]{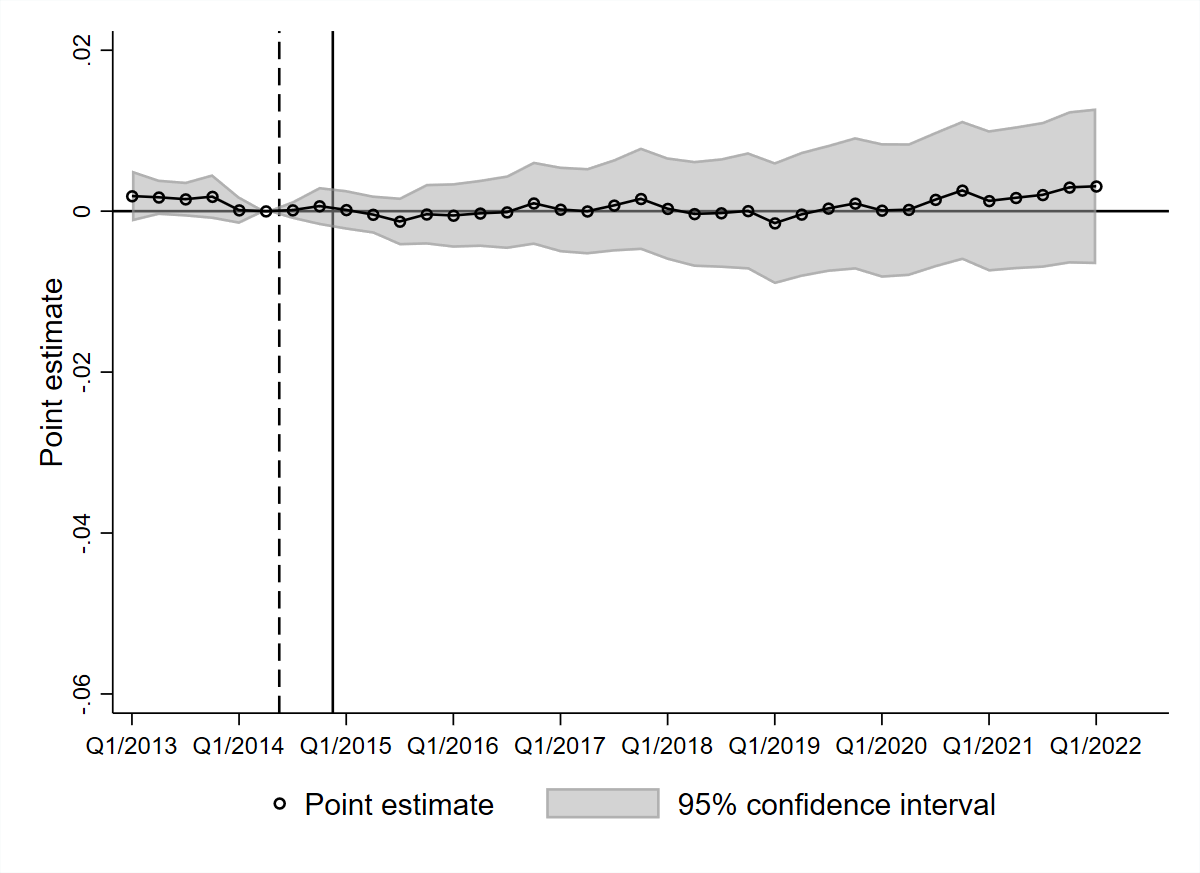}
	\end{subfigure}
 \\
         \center
	\begin{subfigure}{0.5\textwidth} 
		\subcaption{Marginal employment} \label{fig:geb_main}\includegraphics[scale=0.18, trim=0 0 0 0, clip]{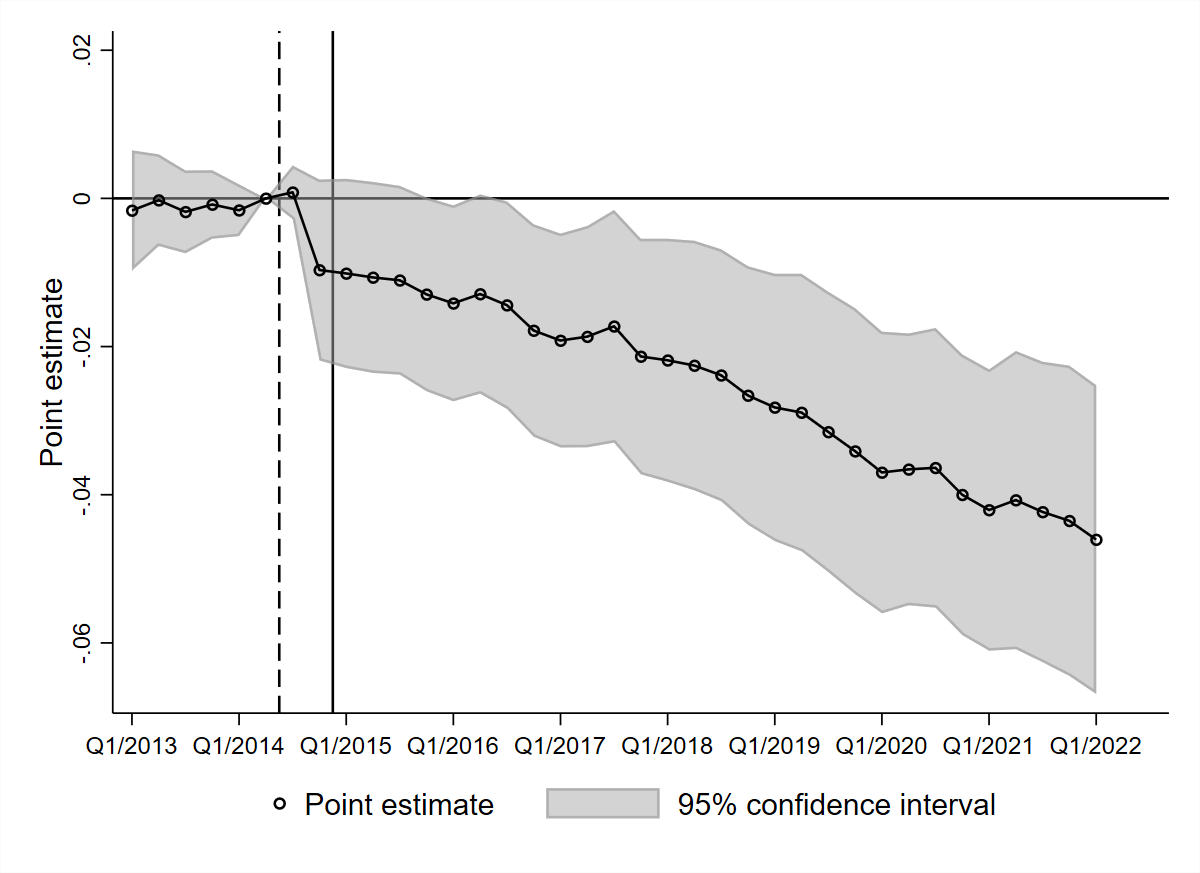}
	\end{subfigure}

	\caption*{\scriptsize Source: Regional Statistic of the Federal Employment Agency, SES 2014, Federal Statistical Office, Federal Institute for Research on Building, Urban Affairs and Spatial Development (FBUS); own calculations. Note: The vertical lines show the points of time when the minimum wage law was passed (August 2014) and introduced (January 1, 2015). The point estimates and confidence intervals refer to the vector $\beta$ in equation \ref{eq:equation2}. The control variables included are interactions of: time, east, and population share between 18 and 64 years in 2013; time, east, and type of labour market region; time, east, and employment share in different sectors; time and GDP per capita in 2013. Standard errors (in parentheses) are clustered at the level of the labour market regions.}
\end{figure}

Figure \ref{fig:event_main} depicts the results of the extended DiD approach of Equation (\ref{eq:equation2}). In addition to the point estimator for each quarter, it also shows the 95-percent confidence intervals as well as vertical lines that mark the announcement of the minimum wage and its introduction. Again, the specifications include all of the time-variant control variables. The assumption of parallel pre-trends seems to hold for all three outcomes, since the point estimates prior to the second quarter of 2014 are not significantly different from zero. Similar to the simple DiD model, the results show a continuously negative effect on the total dependent employment. \linebreak

\noindent The coefficients switch between significance at the 5 and 10 percent level throughout the observation period. This negative effect is hereby fully attributable to the marginal employment for which the effect strongly increased over time in absolute terms. There is no statistically significant effect on the employment subject to social security contributions, although there appears to be a  positive (albeit not significant) trend over the last few quarters during the observation period.

\paragraph{Interaction with Economic Growth}

In order to ascertain whether regions are affected differently by the treatment based on their economic growth prior to the introduction of the minimum wage, we interact the binary treatment with an indicator for low growth between 2010 and 2013. Hereby, low growth is defined as a growth rate in the bottom quartile among all labour market regions. 

The results for the total effect are presented in Panel B of Table \ref{tab:results_main_bip}. The main treatment coefficient for dependent employment remains negative but is no longer significant. The estimate for the interaction with a low growth rate is negative but also insignificant. For employment subject to social security contributions, the main treatment indicator is again positive and insignificant, while the interaction term with economic growth is also negative and insignificant. The main treatment effect for marginal employment is statistically significant at the 5-percent level and negative, but has decreased in magnitude to about $-1.9$ percent. The interaction with a low growth rate prior to treatment start is significant at the 10-percent level at $-2.3$ percent. While the marginal employment decreases in all treated regions on average, it does so more than twice as much in regions with a low growth dynamic between 2010 and 2013. 

\begin{figure}[h!] \caption{Effect of the minimum wage on employment interacted with the regional economic growth} \label{fig:event_growth}
	\begin{subfigure}{0.5\textwidth}
		\subcaption{Dependent employment} \label{fig:svgeb_growth}\includegraphics[scale=0.18, trim=0 0 0 0, clip]{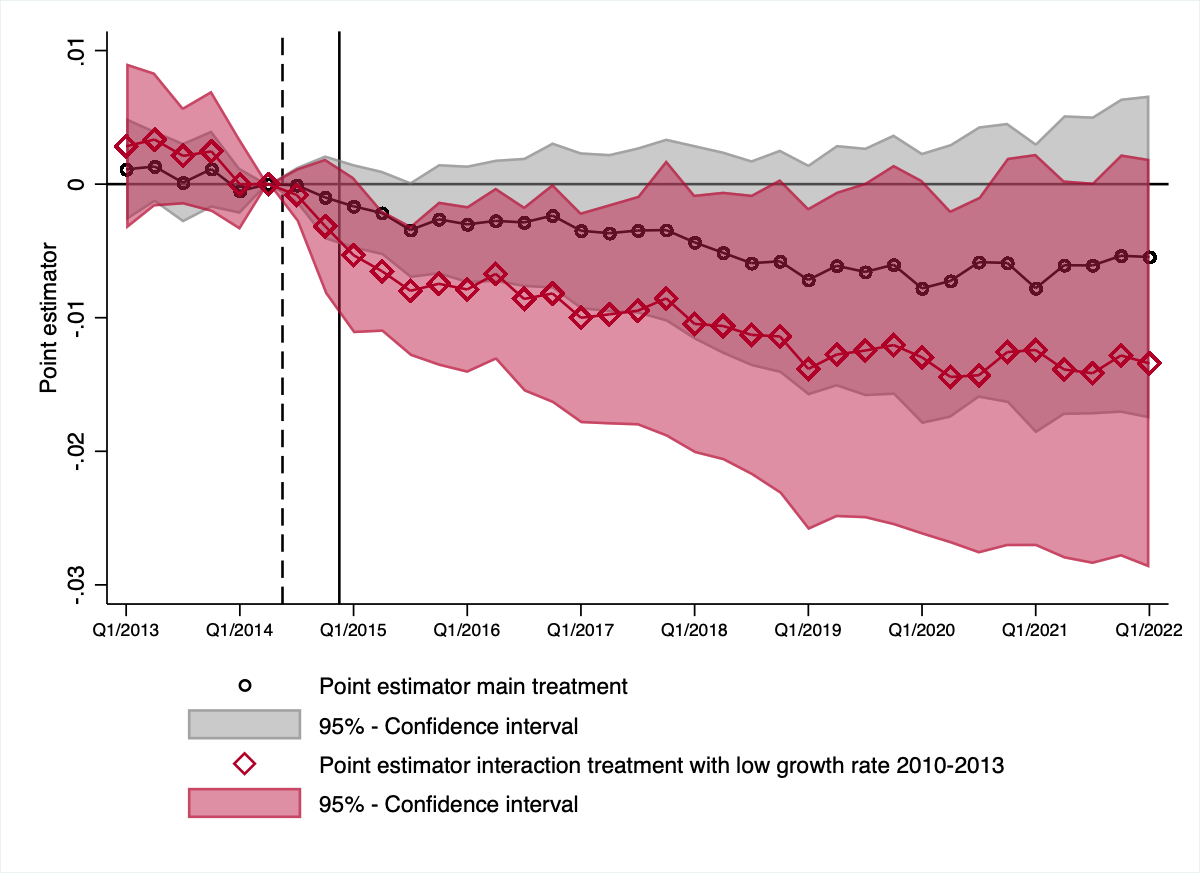}
	\end{subfigure}
	\begin{subfigure}{0.5\textwidth} 
		\subcaption{Employment subject to SSC} \label{fig:svb_growth}\includegraphics[scale=0.18, trim=0 0 0 0, clip]{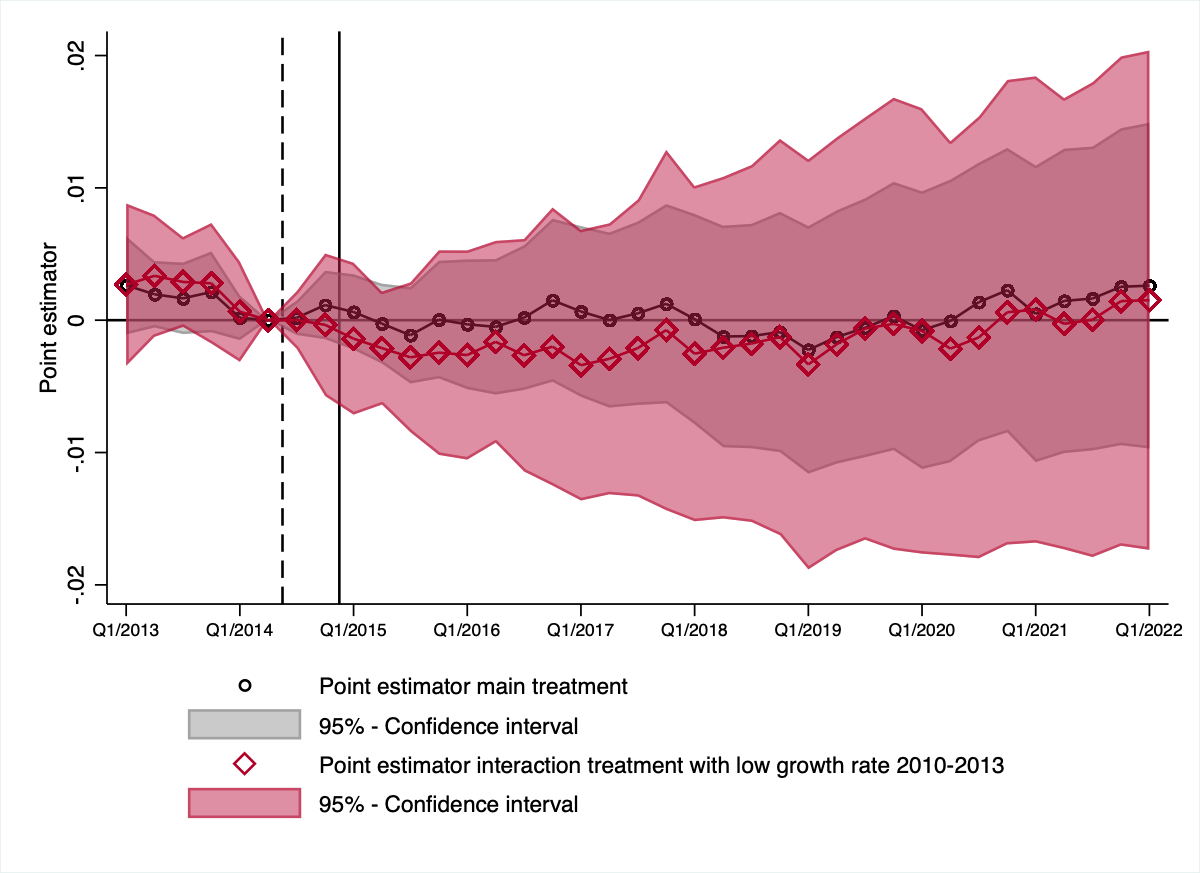}
	\end{subfigure}
 \\
         \center
	\begin{subfigure}{0.5\textwidth} 
		\subcaption{Marginal employment} \label{fig:geb_growth}\includegraphics[scale=0.18, trim=0 0 0 0, clip]{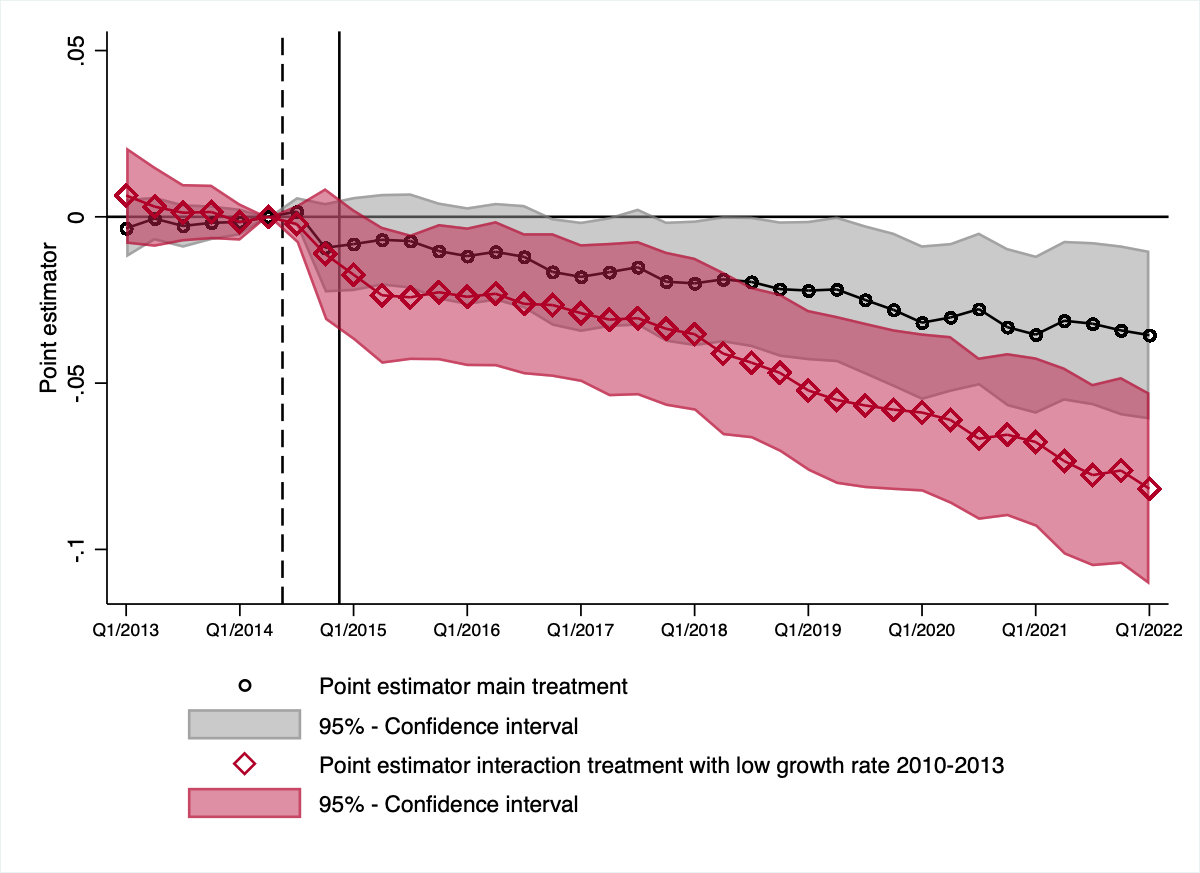}
	\end{subfigure}

	\caption*{\scriptsize Source: Regional Statistic of the Federal Employment Agency, SES 2014, Federal Statistical Office, Federal Institute for Research on Building, Urban Affairs and Spatial Development (FBUS); own calculations. Note: The vertical lines show the points of time when the minimum wage law was passed (August 2014) and introduced (January 1, 2015). 
    The point estimates and confidence intervals refer to coefficients of Equation (\ref{eq:equation2}) augmented by an interaction of the binary treatment and a binary indicator for having a regional GDP growth in the lower 25\% of all regions and an interaction of time and low growth between 2010 and 2013. The black markers show the point estimates for the main treatment indicator and the red markers those for the interaction of treatment and low growth between 2010 and 2013. The control variables included are interactions of: time, east, and population share between 18 and 64 years in 2013; time, east, and type of labour market region; time, east, and employment share in different sectors; time and GDP per capita in 2013. Standard errors (in parentheses) are clustered at the level at the labour market regions.}
\end{figure}

Figure \ref{fig:event_growth} shows the event-study results separate for the treatment interacted with a relatively low growth rate and the main treatment indicator (which shows the treatment effect for regions with a relatively high growth rate). In Panels (a) and (b), it becomes clear that the treatment effects on dependent employment and employment subject to social security contributions for regions with a relatively high growth rate are insignificant throughout the observation period. By contrast, the effect on dependent employment among regions with a relatively low GDP growth is significantly negative (at the 5 percent level) in the majority of periods. For marginal employment, the main treatment effects turn significantly negative at the end of 2018, while they are negative and significant right from the start for regions with a low economic dynamic.

\clearpage

\subsection{Effects of the Minimum Wage Increases}\label{subsec:results_increases}

 Thus far, we have presented estimates of the longer-term effects of the first-time introduction of the German minimum wage in 2015. In the next step, we explicitly consider the subsequent increases in the level of the minimum wage after its introduction. For this purpose, we apply Equation (\ref{eq:equation3}), which includes interaction terms of the treatment and indicators for the post-increase periods. We include the minimum wage increases from January 1 of each year between 2017 and 2022 (except 2018). We do not consider the increase from July 1, 2021, due to its small magnitude. We also consider a shorter version that only includes the increases in 2017 and 2019. The results are presented in Table \ref{tab:increases_VSE2014_bip}, whereby Panel A presents the results for the short estimation with only two increases and Panel B those for the entire Equation (\ref{eq:equation3}). 
 
 The estimate for the treatment effect of the minimum wage introduction for dependent employment in column (1) shows a significant effect (at the 5 percent level) of $-0.3$ percent in both Panels A and B, while the placebo terms are insignificant. The effect size is smaller in magnitude compared to the results for the introduction alone in Table \ref{tab:results_main_bip}. The coefficients for the increases between 2017 and 2020 are negative and turn positive for the subsequent increases. However, they are all very small and insignificant, suggesting that the minimum wage increases had no additional significant effects on the overall dependent employment. The main treatment effect on employment subject to social security contributions in column (2) is insignificant in both specifications, as was the case in the main results in Table \ref{tab:results_main_bip}. The coefficients for all raises are insignificant and with the exception of the estimate for the increase in 2019 in Panel B, they are all positive, yet very small. Finally, the main treatment coefficient for marginal employment is similar and significant at the 10 percent level in both panels with $-1.1$ percent. Thus, it is smaller in magnitude compared to Table \ref{tab:results_main_bip}. In addition to the introduction, it appears that all increases had significant additional negative effects between 0.4 and 1.6 percent. Overall, these results suggest significant additional effects of the increases of the minimum wage, especially for the first two in 2017 and 2019 and on marginal employment.

 \begin{scriptsize} 
	\begin{table}[h!]
  \centering
  \caption{Effects of the minimum wage introduction and its raises on regional employment}
  \label{tab:increases_VSE2014_bip}
    \begin{tabular}{lccc}
    \toprule
    \toprule
    \multicolumn{1}{l}{} & (1)   & (2)  & (3) \\
    \multicolumn{1}{l}{VARIABLES} & Dependent  & Employment  & Marginal  \\
   & employment  & subject to SSC  & employment  \\
\midrule
\multicolumn{4}{l}{\textbf{Panel A: Minimum Wage increases in 2017 and 2019}} \\
          &       &       &  \\
    \multicolumn{1}{l}{Treatment 2014   } & -0.00294** & -0.000116 & -0.0113* \\
    \multicolumn{1}{l}{} & (0.00144) & (0.00151) & (0.00590) \\
    \multicolumn{1}{l}{Placebo} & 0.000595 & 0.00141 & -0.00124 \\
    \multicolumn{1}{l}{} & (0.00112) & (0.00107) & (0.00246) \\
    \multicolumn{1}{l}{Treatment 2014 x (Time $>$ 2016)} & -0.00197 & 0.000380 & -0.0101*** \\
    \multicolumn{1}{l}{} & (0.00188) & (0.00194) & (0.00357) \\
    \multicolumn{1}{l}{Treatment 2014 x (Time $>$ 2018)} & -0.00233 & 0.000845 & -0.0160*** \\
          & (0.00188) & (0.00188) & (0.00395) \\
          &       &       &  \\
    \multicolumn{1}{l}{R$^{2}$ (within)} & 0.598 & 0.585 & 0.487 \\
          &       &       &  \\
    \multicolumn{4}{l}{\textbf{Panel B: Minimum Wage increases in 2017, 2019, 2020, 2021, and 2022}} \\
          &       &       &  \\
    \multicolumn{1}{l}{Treatment 2014   } & -0.00294** & -0.000111 & -0.0113* \\
    \multicolumn{1}{l}{} & (0.00144) & (0.00151) & (0.00590) \\
    \multicolumn{1}{l}{Placebo} & 0.000595 & 0.00141 & -0.00123 \\
    \multicolumn{1}{l}{} & (0.00112) & (0.00107) & (0.00246) \\
    \multicolumn{1}{l}{Treatment 2014 x (Time $>$ 2016)} & -0.00197 & 0.000388 & -0.0101*** \\
    \multicolumn{1}{l}{} & (0.00188) & (0.00194) & (0.00357) \\
    \multicolumn{1}{l}{Treatment 2014 x (Time $>$ 2018)} & -0.00219 & -0.000423 & -0.00925*** \\
    \multicolumn{1}{l}{} & (0.00133) & (0.00130) & (0.00310) \\
    \multicolumn{1}{l}{Treatment 2014 x (Time $>$ 2019)} & -0.000443 & 0.00122 & -0.00678*** \\
    \multicolumn{1}{l}{} & (0.00114) & (0.00119) & (0.00239) \\
    \multicolumn{1}{l}{Treatment 2014 x (Time $>$ 2020)} & 0.000305 & 0.000907 & -0.00466** \\
    \multicolumn{1}{l}{} & (0.000976) & (0.00105) & (0.00230) \\
    \multicolumn{1}{l}{Treatment 2014 x (Time $>$ 2021)} & 0.000650 & 0.00111 & -0.00389** \\
    \multicolumn{1}{l}{} & (0.000727) & (0.000747) & (0.00174) \\
          &       &       &  \\
    \multicolumn{1}{l}{R$^{2}$ (within)} & 0.598 & 0.585 & 0.488 \\
    \multicolumn{1}{l}{} &       &       &  \\
    \midrule
    \multicolumn{1}{l}{Observations} & 9,509 & 9,509 & 9,509 \\
    \midrule
    \multicolumn{1}{l}{Labour market region FE} & X     & X     & X \\
    \multicolumn{1}{l}{Quarter FE} & X     & X     & X \\
    \multicolumn{1}{l}{Controls} & X     & X     & X \\
    \bottomrule
    \bottomrule
    \end{tabular}%
 \caption*{\footnotesize Source: Regional Statistic of the Federal Employment Agency; SES 2014; Federal Statistical Office; Federal Institute for Research on Building, Urban Affairs and Spatial Development; own calculations. Note: The treatment effects refer to the coefficients $\delta_{\tau}$ in Equation (\ref{eq:equation3}), estimated with TWFE. The binary treatment equals 1 if the regional wage gap is equal to or above the population-weighted median. The control variables included are interactions of: time, east, and population share between 18 and 64 years in 2013; time, east, and type of labour market region; time, east, and employment share in different sectors; time and GDP per capita in 2013. Standard errors (in parentheses) are clustered at the level of the labour market regions. Confidence level: ***p$<$0.01, **p$<$0.05, *p$<$0.1.}
\end{table}%
 \end{scriptsize}
 
\clearpage

\subsection{Effects of the Minimum Wage for Switching Treatment Groups}\label{subsec:results_groups}

In Section \ref{sec:data_desc}, we have documented that the regional wage gap substantially changed between 2014 and 2018, as well as the regional ranking by size of the wage gap. We take these changing wage gaps into account by estimating effects for three distinct treatment groups that are defined based on the relative size of the regional wage gap with respect to the median in 2014 and 2018. As already described in Section \ref{sec:data_desc}, these groups are as follows: low wage gap in 2014, high wage gap in 2018 (low/high group); high wage gap in 2014, low wage gap in 2018 (high/low group); high wage gap in 2014 and 2018 (high/high group). The group of regions with a relatively low wage gap in both years form the control group for this analysis (low/low group). This specification accounts for the whole treatment path combined instead of single treatments at different points in time. In addition to the three treatment indicators, we also add a placebo term for each of them. 

\begin{scriptsize}
\begin{table}[t!]
  \centering
  \caption{Effects of the minimum wage introduction on regional employment with multiple treatment groups }
  \label{tab:groups_binary_bip}
    \begin{tabular}{lccc}
    \toprule
    \toprule
    \multicolumn{1}{l}{} & (1)  & (2)  & (3) \\
    \multicolumn{1}{l}{VARIABLES} & Dependent & Employment  & Marginal  \\
    & employment  & subject to SSC  & employment  \\
    \midrule
    \multicolumn{1}{l}{} &       &       &  \\
      \multicolumn{1}{l}{Group low/low: Reference group} &  &  &  \\
    \multicolumn{1}{l}{} &  &  &  \\
    \multicolumn{1}{l}{Group low/high} & -0.00710 & -0.00511 & -0.0208*** \\
    \multicolumn{1}{l}{} & (0.00439) & (0.00448) & (0.00676) \\
    \multicolumn{1}{l}{Group high/low} & -0.00372 & 0.00165 & -0.0207* \\
    \multicolumn{1}{l}{} & (0.00364) & (0.00411) & (0.0117) \\
    \multicolumn{1}{l}{Group high/high} & -0.0108*** & -0.00378 & -0.0401*** \\
    \multicolumn{1}{l}{} & (0.00286) & (0.00335) & (0.00956) \\
    \multicolumn{1}{l}{Placebo Group low/high} & 0.00181 & 0.00259 & -0.000553 \\
    \multicolumn{1}{l}{} & (0.00177) & (0.00166) & (0.00351) \\
    \multicolumn{1}{l}{Placebo Group high/low} & 0.00167 & 0.00288** & -0.000956 \\
    \multicolumn{1}{l}{} & (0.00128) & (0.00134) & (0.00340) \\
    \multicolumn{1}{l}{Placebo Group high/high} & 0.00108 & 0.00222 & -0.00204 \\
    \multicolumn{1}{l}{} & (0.00144) & (0.00142) & (0.00290) \\
    \midrule
    \multicolumn{1}{l}{Observations} & 9,509 & 9,509 & 9,509 \\
    \multicolumn{1}{l}{R$^{2}$ (within)} & 0.601 & 0.587 & 0.480 \\
    \midrule
    \multicolumn{1}{l}{Labour market region FE} & X     & X     & X \\
    \multicolumn{1}{l}{Quarter FE} & X     & X     & X \\
    \multicolumn{1}{l}{Controls} & X     & X     & X \\
    \bottomrule
    \bottomrule
    \end{tabular}%
    \caption*{\scriptsize Source: Regional Statistic of the Federal Employment Agency; SES 2014; Federal Statistical Office; Federal Institute for Research on Building, Urban Affairs and Spatial Development; own calculations. Note: The treatment effects refer to the coefficients $\beta$, $\gamma$, and $\delta$ in Equation (\ref{eq:equation4}), estimated with TWFE. Group low/low is the reference group and includes regions with a wage gap below the weighted median in 2014 and 2018. Group low/high includes regions with a wage gap below (at or above) the median in 2014 (2018). Group high/low includes regions with a wage gap at or above (below) the median in 2014 (2018). Group high/high includes regions with a wage gap at or above the median in 2014 and 2018. The control variables included are interactions of: time, east, and population share between 18 and 64 years in 2013; time, east, and type of labour market region; time, east, and employment share in different sectors; time and GDP per capita in 2013. Standard errors (in parentheses) are clustered at the level of the labour market regions. Confidence level: ***p$<$0.01, **p$<$0.05, *p$<$0.1.}
\end{table}%

 \end{scriptsize}

The results are presented in Table \ref{tab:groups_binary_bip}. The effect on dependent employment is negative for all three treatment groups and the placebo terms are all insignificant. The estimate is insignificant for both the low/high and high/low groups. The effect is largest and statistically significant at the 1 percent level for the high/high group with $-1.1$ percent. Thus, the effect for the group that is strongly affected by the minimum wage in 2014 and 2018 is about twice as large as the overall effect in Panel A of Table \ref{tab:results_main_bip}. For the employment subject to social security contributions, all three coefficients are insignificant and the placebo term for the high/low group is significant. By contrast, all of the placebo terms are insignificant for the estimation of effects on marginal employment. The coefficients for the low/high and high/low groups are about the same size with almost $-2.1$ percent, respectively. The former is significant at the 1 percent level and the latter at the 10 percent level. With $-4.0$ percent, the effect on marginal employment for the high/high group is almost twice as strong and significant at the 1 percent level.

\subsection{Staggered Treatment Adoption} \label{subsec:results_staggered} 

Another way to exploit both the wage gaps of 2014 and 2018 for treatment estimation is to combine both into a treatment that is adopted in a staggered manner at two points, namely 2014 and 2019. Hereby, we assume that treatment is absorbing, thus a region that is treated once remains treated until the end of the observation period. First, we estimate the effect of this staggered treatment via the TWFE model applied in the previous sections and apply the decomposition of the treatment coefficient introduced by  \cite{Goodman-Bacon2021}. As described in Section \ref{sec:method}, effect estimates of a staggered treatment can be biased if treatment effects are heterogeneous across time and/or regions. Therefore, we test the robustness of the TWFE estimates against using alternative new estimators by \cite{CallawaySant'Anna2021}, \cite{SunAbraham2021}, \cite{dCh_dH2022b}, and \cite{BorusyakEtAl2021}.

Table \ref{tab:main_wageGap_staggered_BIP} in the Appendix shows the estimates for the staggered treatment using the same TWFE model as before (see Equation (\ref{eq:equation5})). The coefficient for the total dependent employment is identical in direction and significance level and very similar in magnitude to that from Table \ref{tab:results_main_bip}. The estimate for the employment subject to social security contributions changes its sign and is now negative, but it remains small and insignificant. While the estimate for the marginal employment is negative and significant -- as was the case in Table \ref{tab:results_main_bip} -- its magnitude has substantially decreased from 2.4 to 1.4 percent. 

We decompose our TWFE estimations according to \cite{Goodman-Bacon2021} (using the user written Stata command ``bacondecomp'' \citep{bacondecomp}) to identify the different components of the overall DiD estimates and their respective weights.\footnote{The detailed decomposition results are provided by the authors upon request.} These components include 2x2 DiDs comparing the two treatment groups to the never-treated group. Further components are comparisons between the earlier- and later-treated group and a within component that is caused by differences in covariates within groups. The overall DiD estimate is the weighted sum of these terms. The decomposition results show that the effect estimates for the 2x2 DiD between the later-treated and never-treated groups are smaller in absolute terms than the 2x2 DiD estimate for the earlier-treated and never-treated groups. The DiD estimates comparing the two treatment groups with each other are positive in all cases and they have a weight of about one-third. The fact that there is substantial weight on these estimates and they include ``forbidden'' comparisons between earlier- and later-treated units supports the notion of testing the robustness of the results using the aforementioned newly developed estimators.

\begin{figure}[t!] \caption{Effect of the minimum wage on employment with staggered treatment adoption and different estimators} \label{fig:staggered_cumulative}
	\centering
	\includegraphics[scale=0.30]{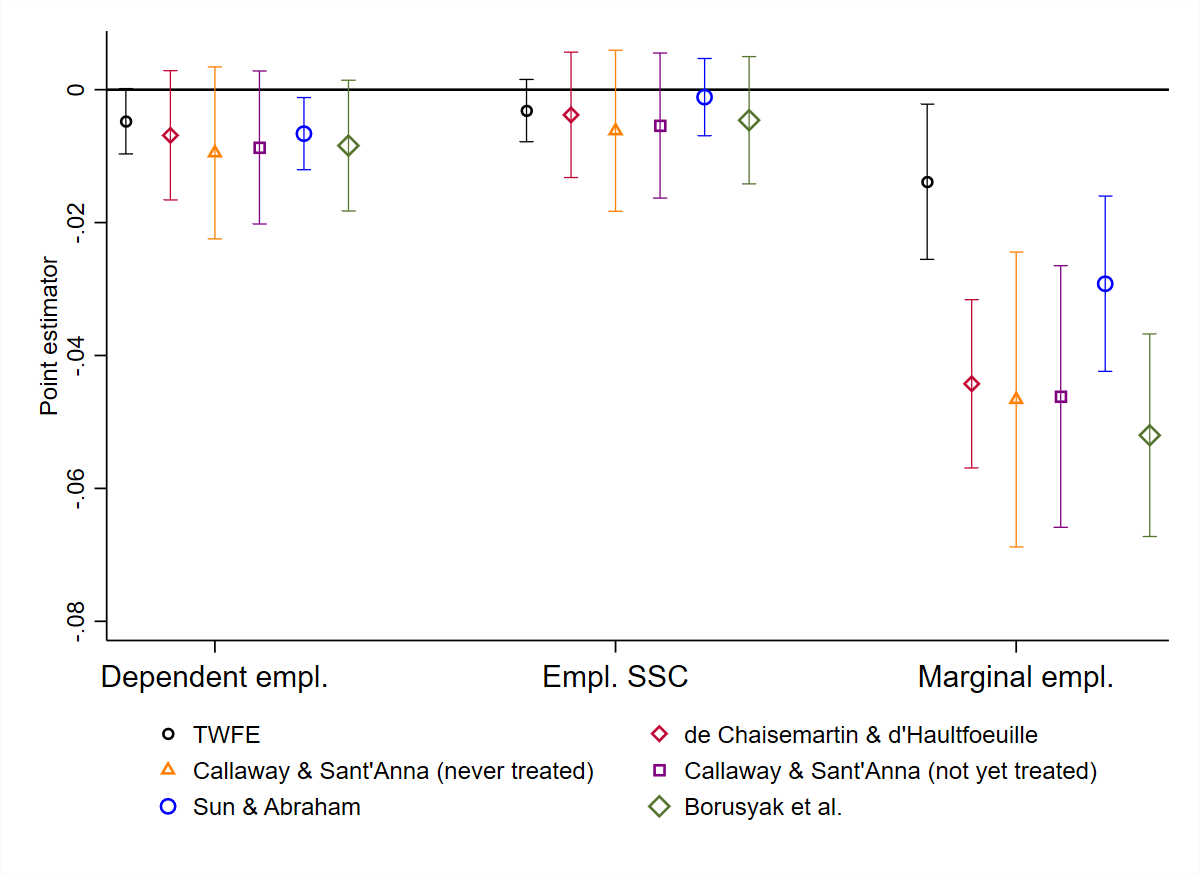}
	\caption*{\scriptsize Source: Regional Statistic of the Federal Employment Agency; SES 2014, 2018; Federal Statistical Office; Federal Institute for Research on Building, Urban Affairs and Spatial Development; own calculations. Note: The treatment effect refers to the coefficient $\beta$ in Equation (\ref{eq:equation5}). The binary treatment equals 1 if the regional wage gap is equal to or above the population-weighted median. The exact choice of control variables varies between estimators due to their varying ability to accommodate covariates. TWFE and \cite{SunAbraham2021}: interactions of: time, east, and population share between 18 and 64 years in 2013; time, east, and type of labour market region; time, east, and employment share in different sectors; time and GDP per capita in 2013. \cite{CallawaySant'Anna2021}: type of labour market region, the GDP per capita in 2013, the population share between 18 and 64 years in 2013, and the employment shares of different sectors in 2013 as constant controls (it can only accommodate invariant controls). \cite{BorusyakEtAl2021}: the same variables as for \cite{CallawaySant'Anna2021}, but interacted with time. \cite{dCh_dH2022b}: reduced number of controls without ``east/west’’ in the interactions.}
\end{figure}

Figure \ref{fig:staggered_cumulative} compares the estimates for the staggered treatment based on the TWFE model with those of the aforementioned estimators by \cite{dCh_dH2022b}, \cite{CallawaySant'Anna2021} (with the never-treated regions and with the not-yet-treated regions as control group), \cite{SunAbraham2021}, and \cite{BorusyakEtAl2021}.\footnote{The results were estimated using the Stata packages ``did\_multiplegt'' \citep{did_multiplegt}, ``csdid'' \citep{csdid}, ``eventstudyinteract'' \citep{eventstudyinteract}, and ``did\_imputation'' \citep{did_imputation}.}$^{,}$\footnote{The exact choice of control variables varies between estimators due to their varying ability to accommodate covariates. The estimator by \cite{SunAbraham2021} allows for the same set of time varying controls as the TWFE estimator described in Section \ref{sec:method}. The estimator by \cite{CallawaySant'Anna2021} can only accommodate invariant controls and thus we add the type of labour market region, the GDP per capita in 2013, the population share between 18 and 64 years in 2013, and the employment shares of different sectors in 2013 as constant controls. For the estimator by \cite{BorusyakEtAl2021}, we include the same variables, but interacted with time. We cannot add ``east/west’’ in the set of interactions because the estimator does not allow controls that perfectly predict treatment status. As all eastern regions start treatment in 2014, this would be the case here. Finally, the estimator by \cite{dCh_dH2022b} appears to have problems with estimation upon adding the large number of interactions included in the TWFE estimation. Thus, we add a reduced number of controls where we leave ``east/west’’ out of the interactions.} All estimates for all three outcomes are negative and thus the direction of the effects is the same, irrespective of using a TWFE or any of the new estimators. The TWFE coefficient for dependent employment is smaller in absolute terms than any of the other estimators, although the difference is only moderate. With the exception of the estimator by \cite{SunAbraham2021}, all of the other coefficients have wider confidence intervals, so that their coefficients are not statistically significant. The TWFE estimate is significant at the 10-percent level and the estimate of \cite{SunAbraham2021} at the 5-percent level. For the employment subject to social security contributions, the effect sizes are even closer to each other, so that no systematic difference exists. None of the coefficients reach any conventional level of significance. The difference between the TWFE and all other coefficients are largest for marginal employment. While all estimates are negative and statistically significant, the effect magnitude is stronger with any of the newly developed estimators compared to the TWFE model. While most of the new estimators yield similar results between $-4.4$ and $-5.2$ percent, the coefficient for \cite{SunAbraham2021} of $-2.9$ percent lies between them and the TWFE estimate with $-1.4$ percent.  

There appear to be two main explanations for the apparent downward bias in the magnitude of the TWFE estimates. First, the treatment effect (compared to the never-treated group) seems to be different for regions starting treatment in 2014 and 2019. As the coefficient is a combination of the treatment effect of regions starting in 2014 and those starting in 2019, this affects the size of the overall estimate. Second, the late-treated group is compared to the early treated group and vice versa. As the early treated regions already experience a treatment effect before the late-treated regions start treatment, this biases the estimated effect on the late-treated regions.

    \section{Conclusion}\label{sec:conclusion}

Our analysis has shown that the introduction of the statutory minimum wage in Germany in 2015 had had significant negative effects on dependent employment during the period until the first quarter of 2022. However, the effect is rather small with -0.5 percent less dependent employment in regions with a relatively high wage gap compared to those with a relatively low wage gap. Moreover, compared to prior studies \citep{PestelEtAl2020} that analysed the period until the first quarter of 2019, the magnitude of the effect decreased. The results further show that the effect is completely attributable to marginal employment, for which the trend of an increasingly negative effect over time became even stronger. We cannot detect a significant effect on employment subject to social security contributions. Interestingly, the overall effect as well as the estimates during the last quarters of the observation period are positive, but very small and insignificant. Our results suggest that the increases of the minimum wage (especially the first two increases in 2017 and 2019) had additional negative effects, particularly on the marginal employment. 

We have been able to extend the existing literature studying the German minimum wage by using measurements of the regional exposure to the minimum wage at two points in time. Based on this data, we found variation in the minimum wage bite over time, whereby about 32 percent of regions changed their treatment status over time. Eighteen (14) percent of these switching labour market regions were weakly (strongly) affected by the minimum wage in 2014 before its introduction and then strongly (weakly) affected in 2018, prior to the second increase. When accounting for these treatment changes over time, we find that employment effects are strongest for the group of regions that was strongly affected at both points in time. Even early on, these regions experienced a larger decline in marginal employment than the regions that were also strongly affected in 2014, but later changed their treatment status. 

Overall, the results are in line with previous studies and they show that even after up to seven years the minimum wage does not seem to have the large negative employment effects that some had predicted prior to its introduction. However, the negative employment effects are stronger for regions with a relatively low GDP growth rate prior to the introduction of the minimum wage, especially for marginal employment. The analysis based on the multiple treatment groups suggests that the regional labour markets developed quite differently even within treatment and control group of 2014. While all regions saw a strong decline in the average wage gap with regard to the current minimum wage between 2014 and 2018, the size of the decrease (in absolute and relative terms) was rather heterogeneous, whereby many regions switched between the treatment and control group. Finally, defining treatment as staggered and applying newly developed estimators alongside the classic TWFE approach reveals that the latter appears to underestimate the effect of the minimum wage on the regional employment, especially on marginal employment. The coefficient of the TWFE estimator suggests an effect of -1.4 percent, while the new estimators mostly suggest a decrease of more than 4 percent. 

There is a need for further research regarding the longer-term effects of the introduction as well as the short- to medium-term effects of the already implemented and upcoming increases on employment and unemployment. The German minimum wage was substantially increased to \euro{12} from October 1, 2022, onwards, which raises the question of non-linearities in the effects of minimum wages. Similar to the introduction of the minimum wage of \euro{8.50} in 2015, the ad-hoc increase to \euro{12} is also a profound cut in the wage structure for many companies and employees. Since it cannot be ruled out that the effects of the statutory minimum wage are characterised by non-linearities, there could be a tipping point from which a further increase has stronger negative employment effects. Similarly, large increases in the minimum wage could have different effects than a sequence of moderate adjustments \citep[see][]{AhlfeldtEtAl2022}. Such non-linearities have not yet been sufficiently explored. The experience with relatively high minimum wages at the local level in the US (for example, \$15 in Seattle) is not representative, as wage levels in the affected cities are higher and thus the depth of intervention of the minimum wage is lower than compared to the nationwide wage level \citep[see][]{Dube2019}.  

It also holds interest to study the mechanism(s) that drive the differential development of the labour market regions after the introduction of the minimum wage. The open questions concern the causes of the substantial changes to the order of the regional wage gaps between 2014 and 2018 and the reasons why some regions with a high wage gap experience a substantially larger negative employment effect than others. 
    \clearpage

\renewcommand{\baselinestretch}{1.12}
\setlength{\baselineskip}{9pt}
\bibliography{MWP_ref.bib}
\nocite{vse14}
\nocite{vse18}
\nocite{regionalstatistik}


\clearpage
\setcounter{figure}{0}
\renewcommand{\thefigure}{A.\arabic{figure}}
\setcounter{table}{0}
\renewcommand{\thetable}{A.\arabic{table}}

\section*{Appendix}
\begin{figure}[h!] \caption{Wage gap ranks of labour market regions in 2014 and 2018} \label{fig:scatter_ranks}
	\centering
	\begin{subfigure}{0.6\textwidth}
		\centering	
		\subcaption{Wage gap ranks 2014 (to 8.50 Euro) and 2018 (to 9.19 Euro)} \label{fig:scatter_ranks1}\includegraphics[scale=0.16]{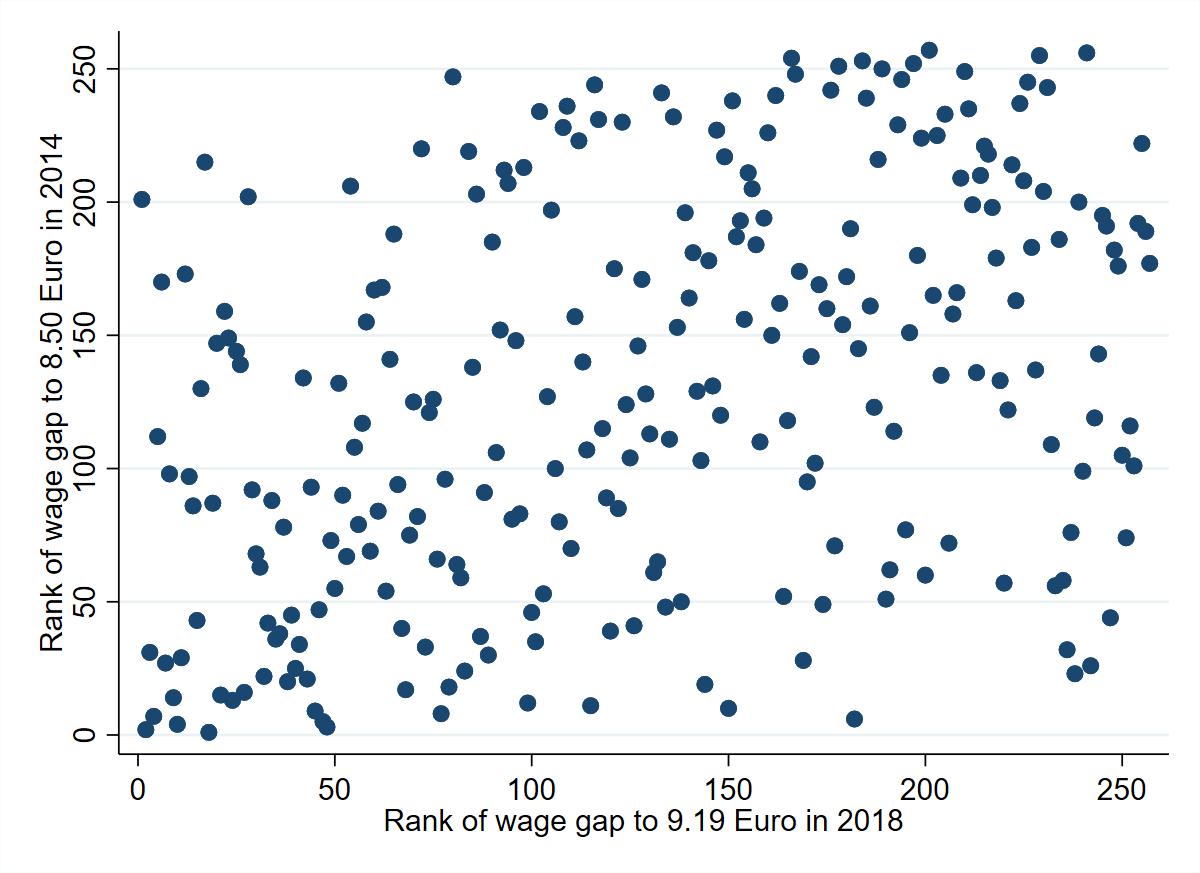}
	\end{subfigure}
	\\
	\begin{subfigure}{0.6\textwidth}
		\centering	
		\subcaption{Wage gap ranks 2014 (to 8.50 Euro and to 9.19 Euro)} \label{fig:scatter_ranks2}\includegraphics[scale=0.16]{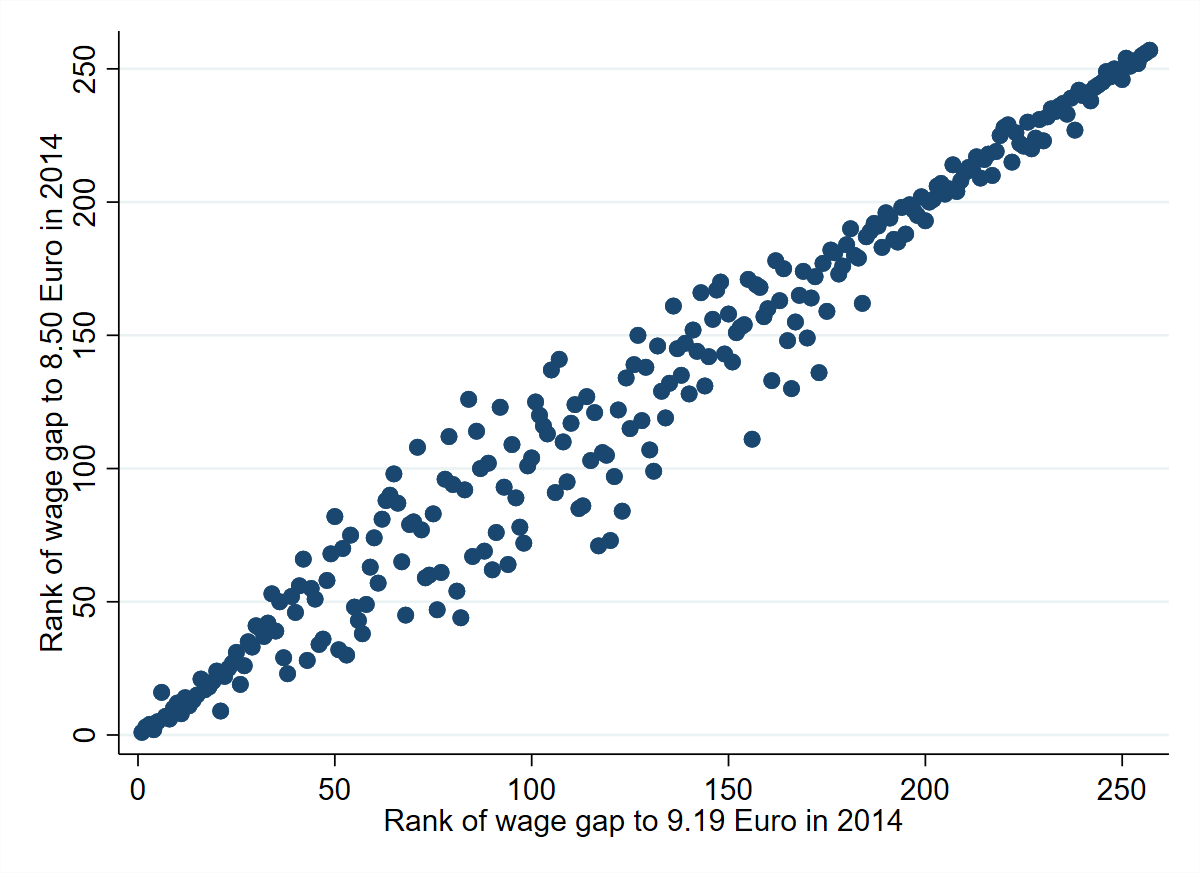}
	\end{subfigure}
	\\
	\begin{subfigure}{0.6\textwidth} 
		\centering
		\subcaption{Wage gap ranks 2014 (to 9.19 Euro) and 2018 (to 9.19 Euro)} \label{fig:scatter_ranks3}\includegraphics[scale=0.16]{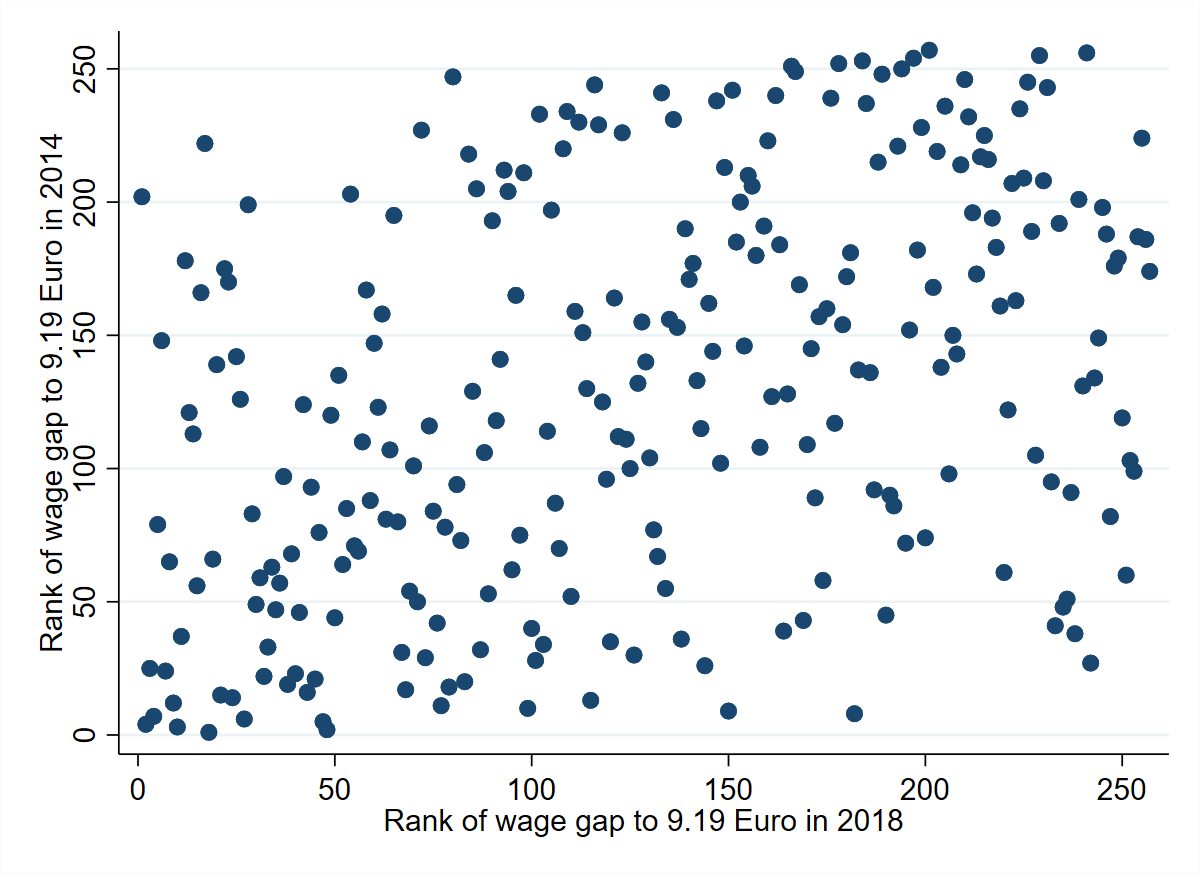}
	\end{subfigure}
	\caption*{\footnotesize Source: SES 2014, 2018; own calculations. Note: Scatter plots of labour market region ranks in terms of the wage gaps in 2014 and 2018. Scatter plot a) plots the ranks of the labour market regions in terms of their wage gap to \euro{8.50} against the rank according to their wage gap to \euro{9.19} in 2018. Scatter plot b) plots the ranks of the labour market regions in terms of their wage gap to \euro{8.50} against the rank according to their wage gap to \euro{9.19} in 2014. Scatter plot c) plots the ranks of the labour market regions in terms of their wage gap to \euro{9.19} against the rank according to their wage gap to \euro{9.19} in 2018.}
\end{figure}

 \newpage

 \begin{figure}[t!] \caption{Outcome evolution over time by treatment status} \label{fig:outcomes}
\center
\includegraphics[scale=0.3]{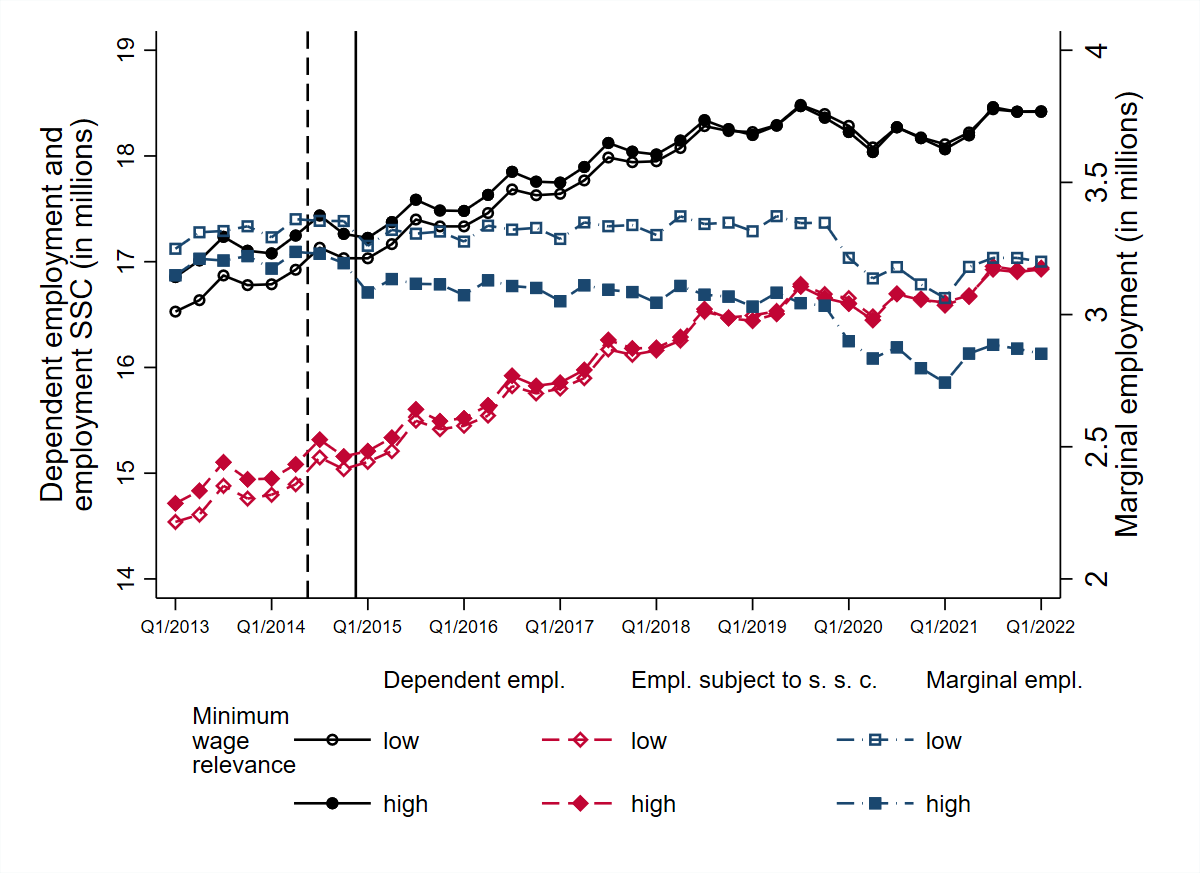}
	\caption*{\footnotesize Source: Regional Statistic of the Federal Employment Agency; SES 2014; own calculations. Note: The graph shows the development of dependent employment, employment subject to social security contributions, and marginal employment between 2013 and 2022 separate for labour market regions with a wage gap at or above and below the population-weighted median.}
\end{figure}

\begin{scriptsize}
	\begin{table}[htbp]
  \centering
\caption{Development of the minimum wage level in Germany}   
\label{tab:mw_development}
\begin{tabular}{lcc}
\toprule
\toprule
			Date & MW level  &  $\Delta$ \\
				 & (in \euro) & (in \%) \\
			\midrule
			 01.01.2015 & 8.50 &  		 	\\
			 01.01.2017 & 8.84 &  4.00		\\
			 01.01.2019 & 9.19 &  3.96		 \\
			 01.01.2020 & 9.35 &  1.74		\\
			 01.01.2021 & 9.50&   1.60		 \\
			 01.07.2021 & 9.60 &  1.05		\\
			 01.01.2022 & 9.82&   2.29		\\
			 01.07.2022 & 10.45 & 6.42	 	\\
			 01.10.2022 & 12.00 & 14.83 	\\
    \bottomrule
    \bottomrule
\end{tabular}\\
 \caption*{\footnotesize Source: Federal Statistical Office; own calculations. Note: The first column shows the date on which the respective increase of the minimum wage was implemented. The second column shows the new level of the minimum wage after it was increased. The third column displays the percentage change of the minimum wage due to the respective increase with respect to the previous level of the minimum wage.}
\end{table}%
 \end{scriptsize}

\begin{scriptsize}
	\begin{table}[htbp]
  \centering
\caption{Descriptive statistics for labour market regions prior to the introduction and the increase of the minimum wage}   
\label{tab:desc_main}
\begin{tabular}{lcccccc}
\toprule
\toprule
&  \multicolumn{3}{c}{SES 2014} & \multicolumn{3}{c}{SES 2018} \\ \cmidrule(lr){2-4}\cmidrule(lr){5-7} 
 Minimum wage exposure & \multirow{2}*{All}  & \multirow{2}*{Low}  &  \multirow{2}*{High} & \multirow{2}*{All}  & \multirow{2}*{Low}  &  \multirow{2}*{High}\\
 (relative to the median wage gap) & & &  & & & \\
\noalign{\smallskip}\hline \noalign{\smallskip} 
 Average wage gap 2014 (in Euro) & 0.203 & 0.104 & 0.281 &  &  &  \\
 Average wage gap 2018 (in Euro) &  &  &  & 0.034 & 0.019 & 0.043 \\
Regions in East Germany in \%) & 21 & 0 & 37.5 & 21 & 7.7 & 30.1 \\
\textbf{Settlement structure (in \%)} &  &  &  &  &  & \\
Urban & 44.7 & 52.2 & 38.9 & 44.7 & 51.9 & 39.9 \\
Rural with tendencies to densification & 24.9 & 22.1 & 27.1 & 24.9 & 25 & 24.8 \\
Sparsely populated, rural & 30.4 & 25.7 & 34 & 30.4 & 23.1 & 35.3 \\
\textbf{Empl. structure by sector 2013 (in \%)} &  &  &  &  &  & \\
Empl. in agriculture, forestry and fishing & 2.4 & 2.3 & 2.5 & 2.4 & 2.1 & 2.6 \\
Empl. in services & 13.7 & 13.7 & 13.7 & 13.7 & 13.9 & 13.5 \\
Empl. in manufacturing & 29.2 & 31 & 27.8 & 29.2 & 30.5 & 28.3 \\
Empl. in the public sector & 30.5 & 28.9 & 31.7 & 30.5 & 29.3 & 31.2 \\
Empl. in trade, transport and hospitality & 24.2 & 24.1 & 24.3 & 24.2 & 24.2 & 24.3 \\
\textbf{Popul. share 18-64 years (2013, in \%)} & 62.4 & 62.6 & 62.2 & 62.4 & 62.6 & 62.2 \\	
\textbf{Economic growth} &  &  &  &  &  & \\
GDP growth rate 2010-2013 (in \%) & 9.8 & 10.2 & 9.5 & 9.8 & 10.4 & 9.4 \\
Low GDP growth 2010-2013 (share in \%) & 25.7 & 30.1 & 22.2 & 25.7 & 25 & 26.1 \\
GDP growth rate 2015-2018 (in \%) & 10.1 & 11.2 & 9.3 & 10.1 & 10.8 & 9.7 \\
Low GDP growth 2015-2018 (share in \%) & 29.2 & 20.4& 36.1 & 29.2 & 22.1 & 34 \\
\midrule
 Number of labour market regions & 257 & 113 & 144 & 257 & 104 & 153 \\
    \bottomrule
    \bottomrule
\end{tabular}\\
 \caption*{\footnotesize Source: SES 2014, 2018; Federal Institute for Research on Building, Urban Affairs and Spatial Development (FBUS), Federal Statistical Office; own calculations. Note: The table presents mean values for all regions and separately by relative size of the regional wage gap for 2014 and 2018. A high (low) minimum wage gap means that the wage gap is above (below or equal to) the median. The division of the labour market regions in types of settlement structure is based on information from the FBUS. The employment by sector, the gross domestic product, and the population shares are taken from the regional statistic from the Federal Statistical Office. Services: Financial, insurance, and business services, real estate and housing. Public services: public and other services, education and health. Trade, transport and hospitality includes information and communication services.}
\end{table}%
 \end{scriptsize}

\begin{scriptsize}
	\begin{table}[htbp]
  \centering
\caption{Descriptive statistics for labour market regions grouped according to their treatment status in 2014 and 2018}  
\label{tab:desc_groups}
\begin{tabular}{lcccccc}
\toprule 
\toprule
 & (1) & (2) & (3) & (4) \\
 Minimum wage exposure in 2014/2018  &Low/ & Low/ & High/ & High/ \\
 (relative to the median wage gap) & Low & High & Low & High  \\
\noalign{\smallskip}\hline \noalign{\smallskip} 
\textbf{Minimum wage bite} &  &  &  &  \\
 Average wage gap 2014 (in Euro) & 0.098 & 0.113 & 0.234 & 0.296  \\
 Average wage gap 2018 (in Euro) & 0.019 & 0.042 & 0.021 & 0.044 \\
 
 \hspace*{5mm}Absolute difference  (in Euro) & -0.079 & -0.071 & -0.214 & -0.252   \\
 \hspace*{5mm}Percentage difference & -80.6 & -62.8 & -91.5 & -85.1 \\
 
 Fraction with hourly wage below  & \multirow{2}*{8.86} & \multirow{2}*{10.35} & \multirow{2}*{16.58} & \multirow{2}*{19.49}  \\
  8.50\euro \hspace*{0.2mm} 2014 (in \%) &  &  &  &   \\
 Fraction with hourly wage below  & \multirow{2}*{5.73} & \multirow{2}*{9.74} & \multirow{2}*{6.22} & \multirow{2}*{10.98}  \\
  9.19\euro \hspace*{0.2mm} 2018 (in \%) &  &  &  &   \\
 
 \hspace*{5mm}Difference in percentage points & -3.13 & -0.61 & -10.36 & -8.51   \\
 \hspace*{5mm}Percentage difference & -35.3 & -5.9 & -62.5 & -43.7 \\
 
\textbf{Regions in East Germany in \%)} & 0 & 0 & 22.2 & 42.6 \\
\textbf{Settlement structure (in \%)} &  &  &  &  \\
Urban & 51.5 & 53.3 & 52.8 & 34.3  \\
Rural with tendencies to densification & 26.5 & 15.6 & 22.2 & 28.7 \\
Sparsely populated, rural  & 22.1 & 31.1 & 25 & 37  \\
\textbf{Empl. structure by sector 2013 (in \%)} &  &  &  &   \\
Empl. in agriculture, forestry and fishing & 2.2 & 2.6 & 2 & 2.7 \\
Empl. in services & 13.9 & 13.4 & 14.1 & 13.6 \\
Empl. in manufacturing & 31.8 & 29.8 & 28 & 27.7 \\
Empl. in the public sector & 28.2 & 30 & 31.5 & 31.8 \\
Empl. in trade, transport and hospitality & 24 & 24.3 & 24.5 & 24.3 \\
\textbf{Popul. share 18-64 years (2013, in \%)} & 62.7 & 62.3 & 62.4 & 62.2 \\	
\textbf{Economic growth} &  &  &  &  \\
GDP growth rate 2010-2013 (in \%) & 10.7 & 9.4 & 9.8 & 9.5 \\
Low GDP growth 2010-2013 (share in \%) & 26.5 & 35.6 & 22.2 & 22.2  \\
GDP growth rate 2015-2018 (in \%) & 11.5 & 10.7 & 9.5 & 9.2 \\
Low GDP growth 2015-2018 (share in \%) & 19.1 & 22.2 & 27.8 & 38.9 \\
\textbf{Hourly wage} &  &  &  & \\
Average hourly wage 2014 (in Euro) & 16.92 & 16.03 & 14.87 & 14.20 \\
Average hourly wage 2018 (in Euro) & 18.67 & 17.32 & 17.50 & 16.02 \\
\hspace*{5mm}Absolute difference  (in Euro) & 1.75 & 1.29 & 2.64 & 1.81   \\
\hspace*{5mm}Percentage difference 2014-2018 & 10.3 & 8.0 & 17.8 & 12.7 \\
\midrule
Number of labour market regions & 68 & 45 & 36 & 108  \\
\bottomrule
\bottomrule 
\end{tabular}\\
 \caption*{\footnotesize Source: SES 2014, 2018; Federal Institute for Research on Building, Urban Affairs and Spatial Development (FBUS), Federal Statistical Office; own calculations. Note: The table presents mean values for four different groups of regions that are defined by the relative size of their wage gaps in 2014 and 2018. A high (low) minimum wage gap means that the wage gap is above (below or equal to) the median. The division of the labour market regions in types of settlement structure is based on information from the FBUS. The employment by sector, the gross domestic product, and the population shares are taken from the regional statistic from the Federal Statistical Office. Services: Financial, insurance, and business services, real estate and housing. Public services: public and other services, education and health. Trade, transport and hospitality includes information and communication services.}
\end{table}%
 \end{scriptsize}

 \begin{scriptsize}
	\begin{table}[htbp]
  \centering
\caption{Effects of the minimum wage introduction on regional employment with a staggered treatment}
  \label{tab:main_wageGap_staggered_BIP}
    \begin{tabular}{lccc}
    \toprule
    \toprule
    \multicolumn{1}{l}{} & (1)   & (2)  & (3) \\
    \multicolumn{1}{l}{VARIABLES} & Dependent & Employment & Marginal   \\
    & employment  & subject to SSC  & employment  \\
    \midrule
    \multicolumn{1}{l}{} &       &       &  \\
    \multicolumn{1}{l}{Treatment (staggered)} & -0.00475* & -0.00315 & -0.0139** \\
          & (0.00251) & (0.00239) & (0.00596) \\
    \multicolumn{1}{l}{} &       &       &  \\
    \midrule
    \multicolumn{1}{l}{Observations} & 9,509 & 9,509 & 9,509 \\
    \multicolumn{1}{l}{R$^{2}$ (within)} & 0.598 & 0.586 & 0.469 \\
    \midrule
    \multicolumn{1}{l}{Labour market region FE} & X     & X     & X \\
    \multicolumn{1}{l}{Quarter FE} & X     & X     & X \\
    \multicolumn{1}{l}{Controls} & X     & X     & X \\
    \bottomrule
    \bottomrule
    \end{tabular}%
  \caption*{\scriptsize Source: Regional Statistic of the Federal Employment Agency; SES 2014; Federal Statistical Office; Federal Institute for Research on Building, Urban Affairs and Spatial Development; own calculations. Note: The treatment effect refers to the coefficient $\beta$ in Equation (\ref{eq:equation5}), estimated with TWFE. The treatment is binary, staggered, and absorbing. Regions with a wage gap equal to or above the population-weighted median in 2014 start treatment in Q3 2014 and regions with a wage gap below (equal to or above) the population-weighted median in 2014 (2018) start treatment in Q1 2019. The control variables included are interactions of: time, east, and population share between 18 and 64 years in 2013; time, east, and type of labour market region; time, east, and employment share in different sectors; time and GDP per capita in 2013. Standard errors (in parentheses) are clustered at the level of the labour market regions. Confidence level: ***p$<$0.01, **p$<$0.05, *p$<$0.1.}
\end{table}%
 \end{scriptsize}

\clearpage
\setcounter{figure}{0}
\renewcommand{\thefigure}{B.\arabic{figure}}
\setcounter{table}{0}
\renewcommand{\thetable}{B.\arabic{table}}

\section*{Supplementary Appendix}
\begin{scriptsize}
	\begin{table}[b!]
  \renewcommand{\arraystretch}{1.5} 
  \centering
  \caption{Results for the Goodman-Bacon decomposition of the TWFE results with a staggered treatment}
  \label{tab:results_bacon_decomp}
    \begin{tabular}{lccc}
    \toprule
    \toprule
    \multirow{2}{2cm}{VARIABLES} & Dependent & Employment  & Marginal  \\
    & employment & subject to SSC  & employment  \\
    \midrule
    Overall DiD estimate & -0.004784 & -0.003177 & -0.013888 \\ 
    Overall DiD variance & 0.000006 & 0.000006 & 0.000036 \\
    \multirow{2}{6cm}{DiD estimate between treatment groups} & \multirow{2}{1.5cm}{0.001759} & \multirow{2}{1.5cm}{0.002070} & \multirow{2}{1.5cm}{0.009555} \\
    & & & \\
    \multirow{2}{6cm}{DiD estimate between early treated and never treated groups} & \multirow{2}{1.6cm}{-0.032767} & \multirow{2}{1.6cm}{-0.022127} & \multirow{2}{1.6cm}{-0.104456} \\
    & & & \\
    \multirow{2}{6cm}{DiD estimate between late treated and never treated groups} & \multirow{2}{1.6cm}{-0.012759} & \multirow{2}{1.6cm}{-0.007378} & \multirow{2}{1.6cm}{-0.031371} \\
    & & & \\
    DiD estimate within  & 0.020987 & 0.011631 & 0.056897  \\
     \multirow{2}{6cm}{Weight on DiD estimate between treatment groups}  & \multirow{2}{1.5cm}{0.354919} & \multirow{2}{1.5cm}{0.354919} & \multirow{2}{1.5cm}{0.354919}  \\
    & & & \\
    \multirow{2}{6cm}{Weight on DiD estimate between early treated and never treated groups} & \multirow{2}{1.5cm}{0.213239} & \multirow{2}{1.5cm}{0.213239} & \multirow{2}{1.5cm}{0.213239} \\
    & & & \\
    \multirow{2}{6cm}{Weight on DiD estimate between late treated and never treated groups} & \multirow{2}{1.5cm}{0.221776} & \multirow{2}{1.5cm}{0.221776} & \multirow{2}{1.5cm}{0.221776} \\
    & & & \\
    Weight on DiD within estimate & 0.210065 & 0.210065 & 0.210065  \\
    \bottomrule
    \bottomrule
    \end{tabular}%
  \caption*{\scriptsize Source: Regional Statistic of the Federal Employment Agency; SES 2014, 2018; Federal Statistical Office; Federal Institute for Research on Building, Urban Affairs and Spatial Development; own calculations. Note: The table shows the results for a decomposition of the TWFE estimate with a staggered treatment (coefficient $\beta$ of Equation (\ref{eq:equation5})) according to \cite{Goodman-Bacon2021}. The coefficients for the overall DiD estimate are negligibly different from those in Table \ref{tab:main_wageGap_staggered_BIP} in the Appendix, because of slight differences in the underlying estimation between the commands ``bacondecomp'' and ``reghdfe'' (the former is used for the decomposition and the latter is used for the TWFE estimations throughout the paper). The regressions underlying the decomposition include all control variables that are also included in all the TWFE estimations throughout the paper. These control variables are interactions of: time, east, and population share between 18 and 64 years in 2013; time, east, and type of labour market region; time, east, and employment share in different sectors; time and GDP per capita in 2013. ``Between treatment groups'' refers to 2x2 DiDs between the later and the earlier treated groups. This includes valid estimates for the treatment effect on the earlier treated group with the later treated group as control group as well as ``forbidden'' comparisons with the earlier treated group as the control group for the later treated group. ``Between early (late) treated and never treated groups'' refers to 2x2 DiDs between the earlier (later) and the never treated groups. ``within estimate'' refers to the component of the overall DiD estimate that is caused by differences in covariates within groups.}
\end{table}%

 \end{scriptsize}


\end{document}